\documentclass[iop]{emulateapj}

\usepackage{graphicx}
\usepackage{amssymb}
\usepackage{amsmath}
\usepackage[breaklinks,colorlinks,
urlcolor=magenta,citecolor=blue,linkcolor=blue]{hyperref}
\usepackage{natbib}
\slugcomment{Accepted for publication in ApJ}

\shorttitle{A Comprehensive Study of L\lowercase{y$\alpha$} Emission in the High-redshift Galaxy Population}
\shortauthors{Oyarz\'un et al.}

\begin{document}


\title{A Comprehensive Study of L\lowercase{y$\alpha$} Emission in the High-redshift Galaxy Population}


\author{Grecco A. Oyarz\'un\altaffilmark{1,2}, Guillermo A. Blanc\altaffilmark{1,3,4}, Valentino Gonz\'alez\altaffilmark{1,4,5}, Mario Mateo\altaffilmark{6}, and John I. Bailey III\altaffilmark{6}
	}

\affil{$^{1}$Departamento de Astronom\'ia, Universidad de Chile, Casilla 36-D, Santiago, Chile; \href{mailto:goyarzun@ucsc.edu}{goyarzun@ucsc.edu}}
\affil{$^{2}$Astronomy Department, University of California, Santa Cruz, CA 95064, USA}
\affil{$^{3}$The Observatories of the Carnegie Institution for Science, 813 Santa Barbara Street, Pasadena, CA 91101, USA}
\affil{$^{4}$Centro de Astrof\'isica y Tecnolog\'ias Afines (CATA), Camino del Observatorio 1515, Las Condes, Santiago, Chile}
\affil{$^{5}$Chinese Academy of Sciences South America Center for Astronomy, China-Chile Joint Center for Astronomy, \\ Camino del Observatorio 1515, Las Condes, Santiago, Chile}
\affil{$^{6}$Department of Astronomy, University of Michigan, Ann Arbor, MI 48109, USA}


\begin{abstract}
We present an exhaustive census of Lyman alpha (Ly$\alpha$) emission in the general galaxy population at $3<z<4.6$. We use the Michigan/Magellan Fiber System (M2FS) spectrograph to study a stellar mass (M$_*$) selected sample of 625 galaxies homogeneously distributed in the range $7.6<\log{\mbox{M$_*$/M$_{\odot}$}}<10.6$.  Our sample is selected from the 3D-HST/CANDELS survey, which provides the complementary data to estimate Ly$\alpha$ equivalent widths ($W_{Ly\alpha}$) and escape fractions ($f_{esc}$) for our galaxies. We find both quantities to anti-correlate with M$_*$, star-formation rate (SFR), UV luminosity, and UV slope ($\beta$). We then model the $W_{Ly\alpha}$ distribution as a function of M$_{UV}$ and $\beta$ using a Bayesian approach. Based on our model and matching the properties of typical Lyman break galaxy (LBG) selections, we conclude that the $W_{Ly\alpha}$ distribution in such samples is heavily dependent on the limiting M$_{UV}$ of the survey. Regarding narrowband surveys, we find their $W_{Ly\alpha}$ selections to bias samples toward low M$_*$, while their line-flux limitations preferentially leave out low-SFR galaxies. We can also use our model to predict the fraction of Ly$\alpha$-emitting LBGs at $4\leqslant z\leqslant 7$. We show that reported drops in the Ly$\alpha$ fraction at $z\geqslant6$, usually attributed to the rapidly increasing neutral gas fraction of the universe, can also be explained by survey M$_{UV}$ incompleteness. This result does not dismiss reionization occurring at $z\sim7$, but highlights that current data is not inconsistent with this process taking place at $z>7$.

\end{abstract}



\keywords{dark ages, reionization, first stars - galaxies: evolution - galaxies: formation - galaxies: high-redshift - galaxies: statistics - ISM: lines and bands}


\section{Introduction}
In hand with observational progress, the understanding of high redshift galaxies has progressed 
immensely in the last few decades. Almost 20 years ago, the only efficient
way of observing these galaxies was by detecting breaks in their 
spectra with broadband photometry. For example, the 
two main classes of rest-frame UV-selected galaxies are Lyman Break Galaxies (LBGs; \citealt{steidel1996,steidel1999,steidel2003,shapley2003,stark2009,bouwens2015}) and Lyman-alpha Emitters (LAEs; \citealt{cowie1996,cowiehu1998,hu1998,kashikawa2006,shimasaku2006,gronwall2007,ouchi2008,blanc2011,sobral2015}). While the former are observed by their Lyman break at 912 \AA, the latter are detected by their Lyman alpha (Ly$\alpha$) emission at 1216 \AA, which is produced in star forming regions and is subject to resonant scattering in the neutral hydrogen of the ISM. After being, at first, used as a galaxy detection method, Ly$\alpha$ emission from galaxies is now widely used to derive a wide variety of information about the physical properties of galaxies
and the universe as a whole.

A lot of effort has been dedicated to study the process of Ly$\alpha$ emission within galaxies. Radiative transfer analysis focuses on the dependence of Ly$\alpha$ escape on ISM kinematics, clumpiness, and outflows (\citealt{dijkstra2012,duval2014,rivera-thorsen2015,gronke2016,gronke2016b}). These studies yield elementary insights that, when combined with observations of the Ly$\alpha$ escape fraction ($f_{esc}$; \citealt{hayes2011,ciardullo2014}), provide a coherent picture for Ly$\alpha$ emission. Still, despite the complex modeling required to explain this particular type of radiation at small scales, Ly$\alpha$ is actively used as a technique for solving cosmological scale paradigms. Among other things, Ly$\alpha$ emission is being used to study the properties of neutral hydrogen in the interstellar, circumgalactic and intergalactic mediums. For instance, the fraction of LAEs as a function of redshift is used as a proxy for the fraction of neutral gas in the IGM (\citealt{stark2011,ono2012,tilvi2014}). If successful, this insight can be used to trace the epoch of reionization, which we have been able to constrain using QSO sightlines (\citealt{fan2006}) and cosmic microwave background (CMB) measurements (\citealt{bennett2013,planck2016a,planck2016b}). Moreover, with the hope of tracing the sources responsible for reionization, Ly$\alpha$ emission is also conveniently used for characterizing luminosity functions (LFs) at high redshift (\citealt{ouchi2008,sobral2009,dressler2015,santos2016}). Nevertheless, the cosmological questions that we are trying to answer by means of this complex emission are not restricted to the reionization of the universe. For instance, Ly$\alpha$ emission can be used to trace the cosmic web (\citealt{gould1996,cantalupo2005,cantalupo2012,cantalupo2014,kollmeier2010}) and young, metal-poor stellar populations in the early universe (\citealt{sobral2015}). Even more, Ly$\alpha$ emission will also be used to study dark energy through baryonic acoustic oscillations in the clustering of LAEs (\citealt{adams2011, blanc2011}).

As evidenced by small-scale studies, Ly$\alpha$ emission is highly stochastic and complex. The observed equivalent width ($W_{Ly\alpha}$) of Ly$\alpha$ lines is highly dependent on neutral gas opacity and kinematics, clumpiness of the ISM, dust distribution, and, therefore, line of sight toward the observer. Furthermore, typical Ly$\alpha$ emitters are extreme objects in terms of their luminosity, metallicity, and emission line diagnostics (\citealt{trainor2016}). As a consequence, if we are to use Ly$\alpha$ emission as a tracer, a careful and thorough approach is required. However, limitations in survey design make such an approach difficult. For instance, spectroscopic studies of UV continuum detected galaxies, typically LBGs (e.g. \citealt{shapley2003,stark2010,ono2012}), reject galaxies with no significant Lyman break and/or red rest-frame UV spectral energy distributions (SEDs), i.e., these studies do not account for passive and heavily extincted galaxies. Observationally, this technique is limited by the M$_{UV}$ sensitivity of the survey, which translates to incompleteness at low-SFR. On the other hand, samples where galaxies are detected directly in Ly$\alpha$ using narrowband imaging (e.g. \citealt{gronwall2007,ouchi2008}) or blind spectroscopy (e.g. \citealt{blanc2011,dressler2011}) select on Ly$\alpha$ equivalent width (W$_{Ly\alpha}$) and emission line flux. Such methodology does not require continuum selection, which allows for line detection in faint objects. However, follow-up rarely allows for more than spectroscopic confirmation of the line, limiting the use of this technique to the measurement of line fluxes, $W_{Ly\alpha}$, and emission profiles. Even more, such surveys can have non-negligible amounts of line interlopers (\citealt{sobral2016}).

Comprehensive studies of Ly$\alpha$ emission in the overall galaxy population are, therefore, of crucial importance. Correlations between galaxy properties and emission are the statistical manifestation of the radiative escape of Ly$\alpha$ photons. Properties such as stellar mass (M$_*$), star formation rate (SFR), UV slope ($\beta$), and merger rate reveal information of how the escape of Ly$\alpha$ radiation is affected by stellar population ages, gas fraction, dust content, and ISM turbulence, respectively. Acknowledgment of such relations is required to draw conclusions from high-redshift Ly$\alpha$ surveys. \citet{oyarzun2016} show how the normalization and e-folding scale of the W$_{Ly\alpha}$ distribution of $3<z<4.6$ galaxies anti-correlate with M$_*$. In other words, higher M$_*$ objects typically have lower W$_{Ly\alpha}$. More massive galaxies typically have lower gas fractions, but higher gas mass (\citealt{tacconi2013,song2014,genzel2015}). More neutral gas contributes to increase the scatter of Ly$\alpha$ photons, spreading this radiation throughout the galaxy and toward the circumgalactic medium. Higher M$_*$ galaxies also have more dust extinction (\citealt{franx2003,daddi2004,daddi2007,forster-schreiber2004,perez-gonzalez2008}), which leads to more Ly$\alpha$ photon absorption. The escape fraction of this radiation is, therefore, severely affected by the stellar mass of galaxies. These results even explain some discrepancies between LBG and narrowband studies of Ly$\alpha$ emission that originated in sample selection effects. This example highlights the need for proper assessment.

In this work, we present a comprehensive analysis of Ly$\alpha$ emission in the $3<z<4.6$ galaxy population. To this end, we further analyze the data introduced by \citet{oyarzun2016}. This sample is composed of 625 galaxies in the M$_*$ range $7.6<\log{(M_*[M_{\odot}])}<10.6$ from the 3D-HST/CANDELS survey (\citealt{grogin2011, koekemoer2011}). Our work is based on spectroscopic observations of the sample using the Michigan/Magellan Fiber System (M2FS; \citealt{mateo2012}), allowing us to measure Ly$\alpha$ fluxes and use 3D-HST/CANDELS ancillary data. In particular, we study Ly$\alpha$ emission dependence on the M$_*$, SFR, UV luminosity, and dust extinction of galaxies. To do so, we introduce a Bayesian approach to properly compare $W_{Ly\alpha}$ distribution models, account for incompleteness, and quantify the significance of observed correlations. We also use our results to simulate high-redshift Ly$\alpha$ emission surveys, allowing us to infer their biases and imprints on the resulting samples. We finally use the correlations we recover to predict the fraction of LAEs as a function of redshift. This semi-analytic approach provides a baseline for comparing observational drops in the fraction of LAEs at high redshift. This work is structured as follows. In Section \ref{2}, we describe our sample and data set. In Section \ref{3}, we describe our line detection and measurement methodologies. We explain our Bayesian $W_{Ly\alpha}$ distribution analysis in Section \ref{4}. We show our results on the $W_{Ly\alpha}$ distribution dependence on different properties and selection techniques in Sections \ref{5} and \ref{6}. We state in Section \ref{7} our inferences on higher redshift Ly$\alpha$ emission. We present our conclusions in Section \ref{8}. Throughout this work, all magnitudes are in the AB system (\citealt{oke1983}). A CDM cosmology with $H_{0}$ = 70 km s$^{-1}$ Mpc$^{-1}$, $\Omega_{m}$ = 0.3, and $\Omega_{\Lambda}$ = 0.7 was assumed whenever needed.

\section{Data Set}
\label{2}
\subsection{Sample Selection}
\label{2.1}
Our sample was initially composed of 629 galaxies in the COSMOS, GOODS-S and UDS fields. Every object was observed under the 3D-HST/CANDELS program, providing HST and \textit{Spitzer} photometry from 3800 \AA \ to 7.9 $\mu m$ (9 bands for COSMOS, 14 for GOODS-S, and 9 for UDS). We construct our sample using 3D-HST outputs (\citealt{skelton2014}). According to these, our 629 photometric redshifts satisfy $3.25<z_{3D-HST}<4.25$ and have a 95\% probability of $2.9<z<4.25$. Every galaxy also complies with a photometric redshift reliability parameter $Q_{z}\leqslant3$ selection to remove catastrophic outliers (\citealt{brammer2008}). In terms of M$_*$, our galaxies are homogeneously distributed in the range $8<\log({\mbox{M}_*[\mbox{M}_{\odot}]})_{3D-HST}<10.4$. These 3D-HST output values are based on the \citet{bruzualcharlot2003} stellar population synthesis model library with a \citet{chabrier2003} IMF and solar metallicity. Exponentially declining star formation histories (SFHs) with a minimum e-folding time of $log_{10}(\tau/yr) = 7$ and a \citet{calzetti2000} dust attenuation law were also assumed for the calculation (\citealt{skelton2014}). We further emphasize that sample selection was performed based on the values of $z_{3D-HST}$ and M$_*$ detailed in this section, whereas all further analysis in this paper is based on our own estimates (Section \ref{2.3}).

\subsection{Data}
\label{2.2}
Spectroscopy of the full sample was conducted at the Magellan Clay 6.5 m telescope during 2014 December and 2015 February. To this end, we used the M2FS, a multi-object fiber-fed spectrograph. The $1\farcs2$ fibers of this instrument allow the observation of 256 targets within 30 arcmin in a single exposure. We are then able to observe 210 targets in each field simultaneously, while using 40 fibers for sky apertures and five for calibration stars. We used this spectrograph in LoRes mode, which features an expected resolution of R=2000 and a continuum sensitivity of V=24 with S/N=5 in 2 hr (\citealt{mateo2012}). The final data set consists of six exposure hours on each of the three fields with an average seeing of $0\farcs6$.

For data reduction, we developed a custom M2FS pipeline. This routine features standard bias subtraction and dark correction. For wavelength calibration, we use HgArNe lamps observed on each night. The wavelength solution is obtained separately for each fiber, with a typical rms uncertainty of 0.03\AA. We further correct the solution for each fiber using the sky lines in the science spectra. For flat-fielding, we use sky-flats obtained during either twilight or dawn. The correction is calculated separately for each fiber, and features illumination, fiber profile, and a correction to account for the shifting of fiber spectra on the detector due to thermal effects. Despite the fact that we are not using dome-flats, we estimate the uncertainties in our flat-fielding to be $<5$\%. 

Sky subtraction is performed using the 40 sky fibers homogeneously distributed over the field of view and across the detector, where fibers are grouped in blocks. We find the sky solution to be more dependent on fiber location in the CCD than in the sky, leading us to perform sky subtraction separately for each frame fiber block. The solution is computed using a non-parametric spline fit, yielding satisfactory sky subtraction for most sky lines. For the bright sky lines that we are not able to properly subtract, we build emission line masks. For consistency, we use the same mask for all spectra, except for particularly noisy fibers. In those rare cases (11), we mask broad sections of the spectrum. Due to fiber malfunction, we could not obtain spectra for 4 of the 629 targets, leaving our final sample in 625 objects.

\begin{figure*}
	\centering
	\includegraphics[width=7in]{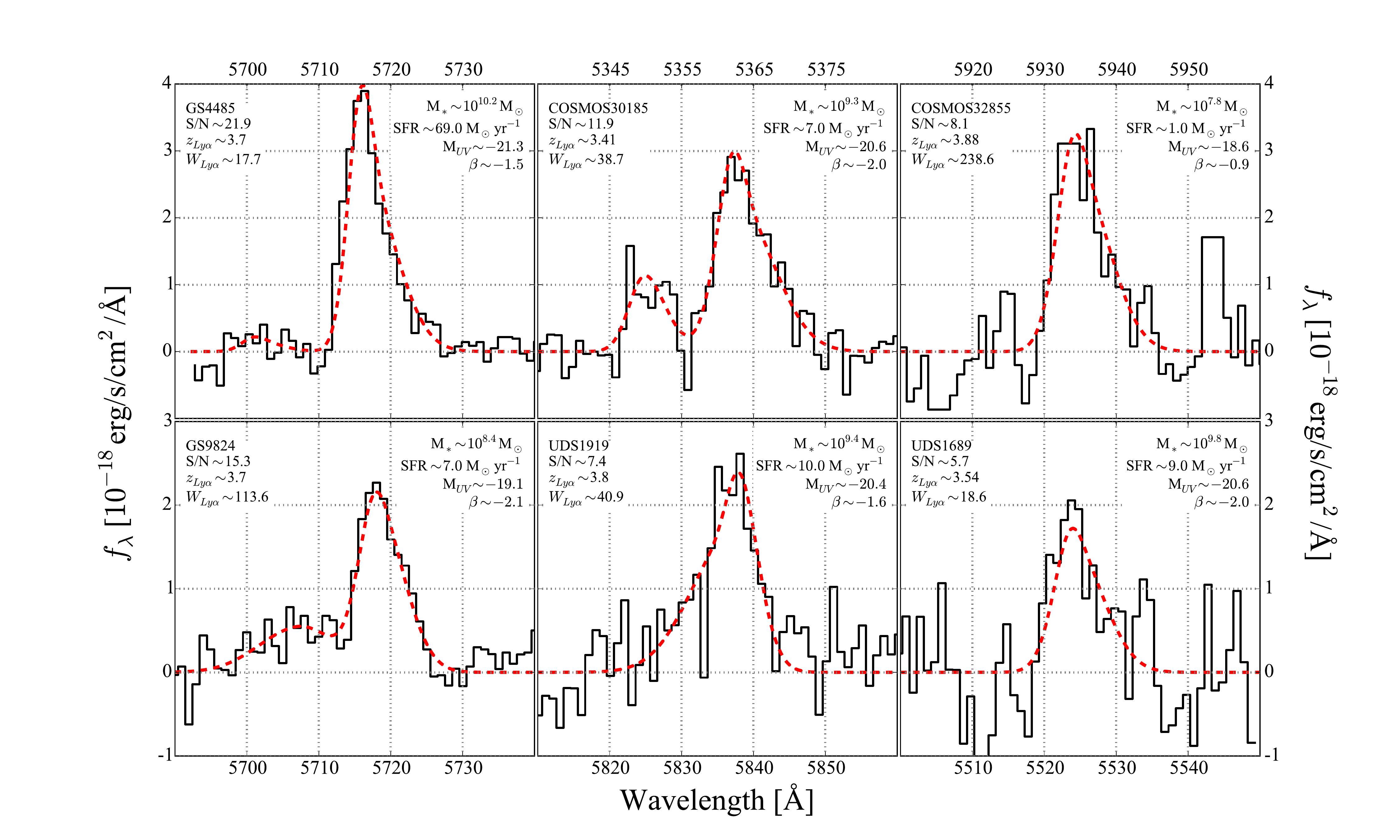}
	\caption{Reduced and continuum subtracted spectra of six Ly$\alpha$ lines (black) from our study. These examples show the wide range of profiles and galaxy properties encompassed by our 120 detections (S/N$\geqslant 5.5$). We also include in these plots the line profiles we fit to each case (red, dashed). The purpose of profile fitting in this work is limited to line-flux measurements.}
	\label{fig:1}	
\end{figure*}

Flux calibration is performed using five M$_V=19-22$ calibration stars on each exposure. Due to atmospheric differential refraction (ADR) comparable to the $1\farcs2$ fiber size, we need to recover the intrinsic spectrum of each star. To do so, we fit stellar templates from the Pickles library (\citealt{pickles1998}) to the continuum-normalized instrumental spectra of the calibration stars. We then use the five stars on each exposure to obtain an average sensitivity curve. We estimate the rms uncertainty of our method to be about $\sim 15\%$. After this calibration, we correct the fluxes in our spectra for Galactic extinction. Samples of sky-subtracted spectra are shown in Figure \ref{fig:1}.  

Considering the three fields, the science area we survey is of $\sim 550$ arcmin$^2$. The resulting spectral FWHM line resolution is of $\sim 2$\AA, and we reach a 1$\sigma$ continuum flux density limit of $\sim 4\times10^{-19}$ erg s$^{-1}$ cm$^{-2}$ \AA$^{-1}$ per pixel in our 6 hr of exposure. We estimate a 5$\sigma$ emission line-flux sensitivity of $\sim 8\times10^{-18}$ erg s$^{-1}$ cm$^{-2}$ in our final spectra (see Figure \ref{fig:1}). We identify 120 Ly$\alpha$ emission lines with S/N$\geqslant 5.5$ in our data (details in Section \ref{3.1}). Given that we have 625 observed targets, then we are then recovering a Ly$\alpha$-emitting galaxy with S/N$\geqslant 5.5$ every $\sim$5 objects.

\begin{figure}
	\centering
	\includegraphics[width=3.4in]{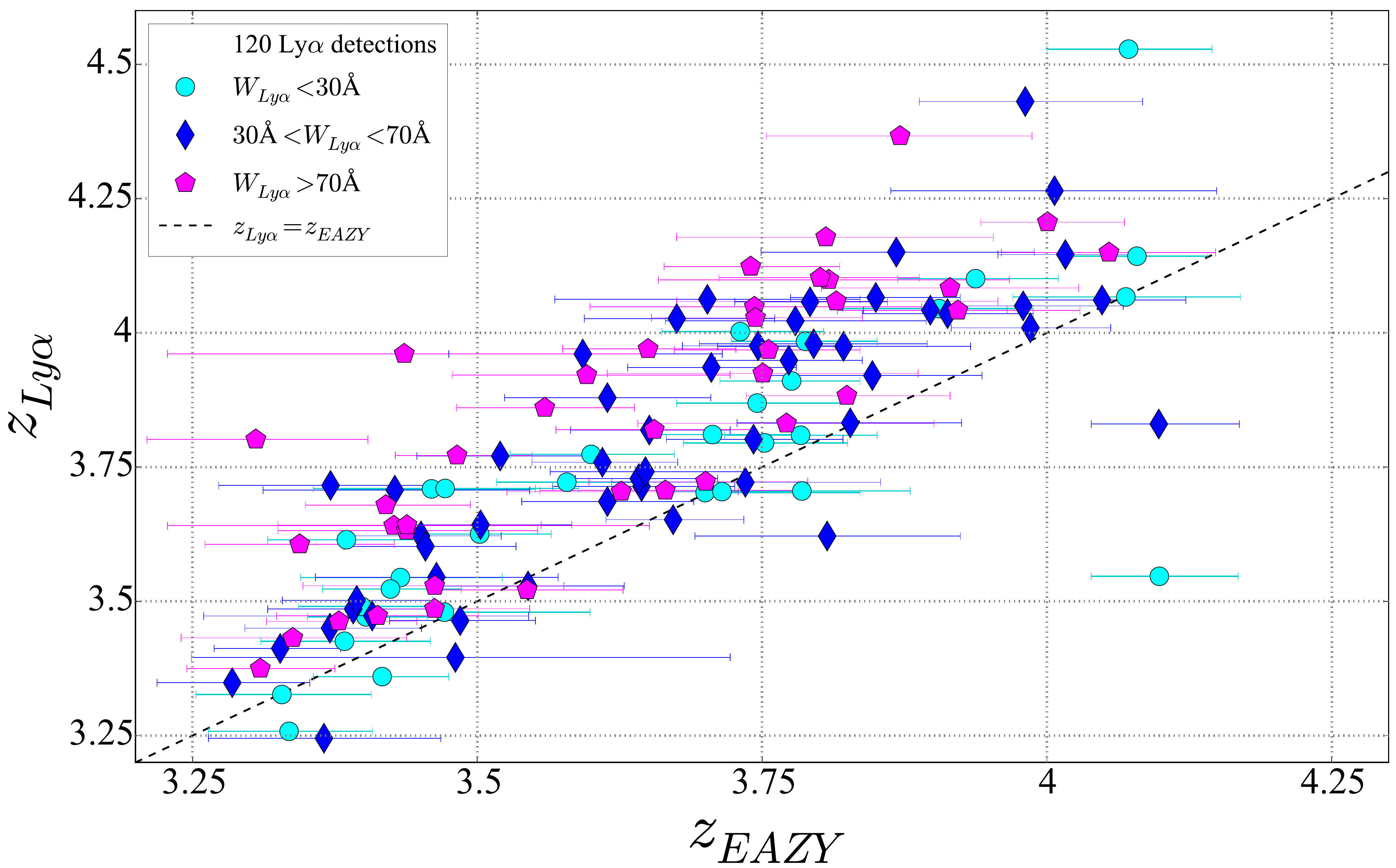}
	\caption{Measured spectroscopic Ly$\alpha$ redshifts ($z_{Ly\alpha}$) as a function of the corresponding photometric counterparts ($z_{EAZY}$) for the 120 Ly$\alpha$ detections (S/N$\geqslant 5.5$). Low-, medium-, and high-$W_{Ly\alpha}$ galaxies are shown as cyan circles, blue diamonds, and magenta pentagons, respectively. We find a median deviation of $\Delta z= z_{Ly\alpha}-z_{EAZY}=0.24$ from the 1:1 relation (dashed). As revealed by this figure, we find the deviation to correlate with $W_{Ly\alpha}$. Hence, we associate this deviation with photometric redshift fitting biases when a strong Ly$\alpha$ emission line is present.}
	\label{fig:2_5}
\end{figure}

\begin{figure}
	\centering
	\includegraphics[width=3.4in]{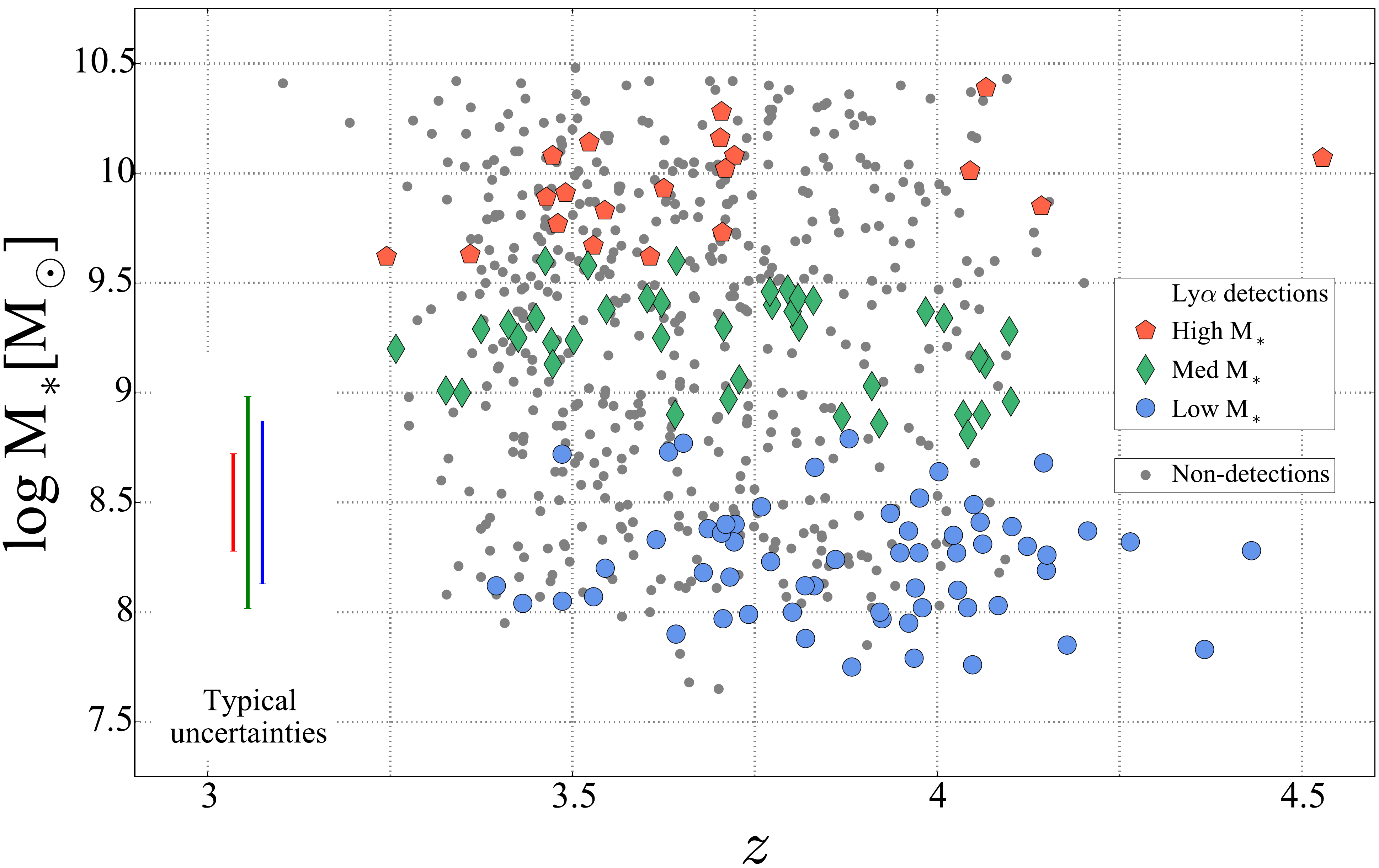}
	\caption{Redshift-M$_*$ distribution of the galaxies composing our sample, according to our executions of EAZY and FAST (Section \ref{2.3}). We divide our sample into three M$_*$ bins using as boundaries log($\mbox{M}_*[\mbox{M}_{\odot}]$)=8.8, 9.6. We find 120 Ly$\alpha$ detections ($S/N\geqslant 5.5$), of which 60 are low M$_*$ (blue circles), 40 are medium M$_*$ (green diamonds), and 20 are high M$_*$ (red pentagons). Non-detections are shown as gray points. For detections, we plot ($z_{Ly\alpha}$, M$_*(z_{Ly\alpha})$), whereas for non-detections we use ($z_{EAZY}$, M$_*(z_{EAZY})$).}
	\label{fig:2}
\end{figure}

\begin{figure}
	\centering
	\includegraphics[width=3.4in]{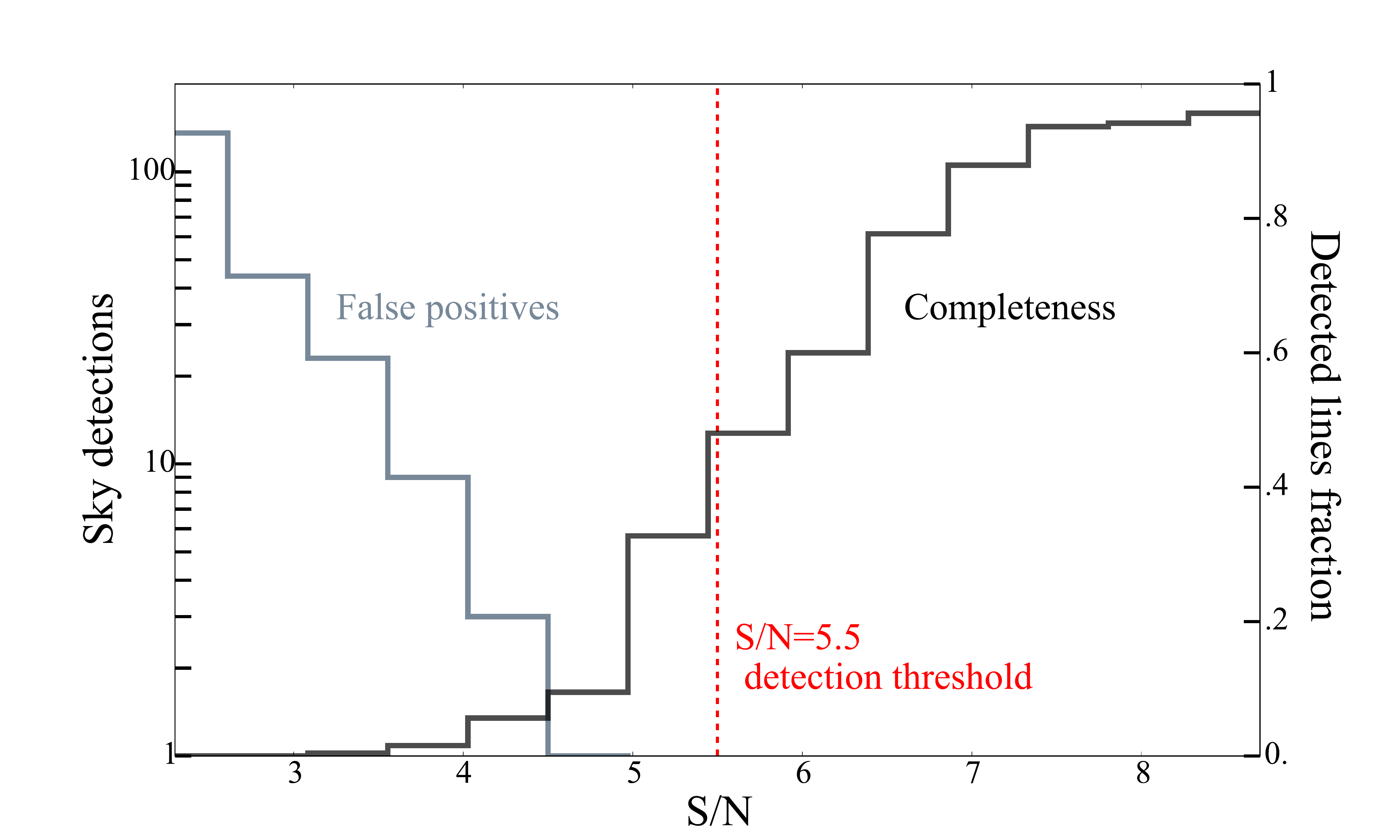}
	\caption{Completeness (black) and false positives (gray) yielded by our line detection method. The completeness histogram is recovered by simulating $\sim 10^3$ Gaussian emission lines on the 115 reduced sky spectra. Hence, for this curve, the x-axis corresponds to S/N$_{sim}$. The false positives histogram is obtained running our line detection code on the 115 skyspectra. The dashed red line shows our line detection threshold S/N=5.5, for which we have a completeness of $\sim 40 \%$.}
	\label{fig:3}
\end{figure}

\subsection{Sample Properties}
\label{2.3}
As stated by \citet{skelton2014}, 3D-HST outputs for these galaxies were obtained using FAST (\citealt{kriek2009}). These calculations assume exponentially declining SFHs with a minimum e-folding time of $log_{10}(\tau/yr) = 7$ (\citealt{skelton2014}). However, since recent studies suggest that it is more adequate to reproduce high-redshift observations with constant SFHs (cSFHs; \citealt{gonzalez2014}), or even rising SFHs (\citealt{maraston2010}), we perform our own executions of FAST assuming cSFHs. For a detailed discussion on this topic, we refer the reader to \citet{conroy2013}. Our FAST executions also adopt the \citet{bruzualcharlot2003} stellar population synthesis model library, a \citet{chabrier2003} IMF, and solar metallicity, similarly to \citet{skelton2014}. We do not account for nebular emission lines in our SED fitting, which, in principle, can overestimate our reported M$_*$ by a factor of up to 4 (\citealt{atek2011,conroy2013,stark2013}) or even $\sim7$ for strong emitters (\citealt{deBarros2014}). However, our galaxies are at $z\sim4$. At this redshift, the overestimate is within a factor of $\lesssim 1.1$ (\citealt{stark2013,salmon2015}), since the 4.5$\mu m$ \textit{Spitzer}/IRAC band is unaffected\footnote{For the redshift range of our galaxies, the H$\beta$, O \scriptsize{II} \footnotesize, and O \scriptsize{III} \footnotesize emission lines fall between the $H_{160}$ and 3.6$\mu m$ bands. H$\alpha$ only contributes to the 3.6$\mu m$ \textit{Spitzer}/IRAC band for $z>3.75$.}. Our FAST outputs also include extinction-corrected SFRs. The SFRs are derived from the SED, and the extinction correction assumes a \citet{calzetti2000} attenuation law with $R_{V}=4.05$.

We are also required to run EAZY (\citealt{brammer2008}) on the photometry from CANDELS/IRAC, since FAST requires redshifts as input. The executions of EAZY yield a most probable redshift $z_{EAZY}$. Our 625 objects satisfy $3<z_{EAZY}<4.25$, with a median uncertainty $\sigma_{EAZY}=0.1$ (Figure \ref{fig:2_5}). EAZY outputs also include $2\sigma$ constraints, which for our sample are limited to $2.95<z<4.5$. Our FAST executions yield a mass coverage of $7.6<\log{(\mbox{M}_*[\mbox{M}_{\odot}])}<10.6$ (Figure \ref{fig:2}), with a characteristic uncertainty of $\log{(\mbox{M}_*[\mbox{M}_{\odot}])}\sim 0.3$. Most of our SFR values are in the range $1-100\mbox{M}_{\odot}/yr$ (further analysis and plots below). From now on, we use the FAST outputs based on the spectroscopic redshifts ($z_{Ly\alpha}$) for the 120 detections (S/N$\geqslant 5.5$; Section \ref{3.1}) and the outputs based on $z_{EAZY}$ for the 505 non-detections.

\subsection{Considerations}
\label{2.4}
The flux calibration procedure performed in our spectra is based on using stars; therefore, it corrects for a $\sim32$\% fiber flux loss, which corresponds to a point-source Ly$\alpha$ surface brightness distribution. In cases of extended Ly$\alpha$ emission halos (\citealt{steidel2011,matsuda2012,feldmeier2013,hayes2013,momose2014,caminha2016,patricio2016,wisotzki2016}), the fluxes we derive are mostly associated with the galaxies themselves and the inner parts of their Ly$\alpha$ halos. Furthermore, Ly$\alpha$ emission from a galaxy can show significant misalignment with the UV continuum (\citealt{rauch2011}), although such cases are not the norm (\citealt{finkelstein2011,jiang2013b}). At the cosmic time of our sample, the fiber diameter corresponds to a scale of $\sim8$ kpc, roughly a factor of four larger than the typical effective diameter of galaxies (\citealt{bond2012,law2012}).

We find a median redshift offset of $\Delta z= z_{Ly\alpha}-z_{EAZY}=0.24$ in our detections (see Figure \ref{fig:2_5}). We show in \citet{oyarzun2016} that this offset does not have any noticeable effects on our sample dependence on M$_*$, which is the primary selection criterion for our galaxies. A very similar offset has also been found in the MUSE-Wide Survey (see \citealt{herenz2017}). We find the offset to correlate with $W_{Ly\alpha}$ (see Figure \ref{fig:2_5}), hinting at biases in photometric redshift fitting when a strong Ly$\alpha$ emission line is present. A plausible scenario is one in which the Lyman break from the template is fitted slightly blueshifted to account for the flux excess in the redder band. Thorough simulations beyond the scope of this work are required to explore the causes behind this bias.

We stress that the 3D-HST/CANDELS mass incompleteness is restricted to $\log{(\mbox{M}_*[\mbox{M}_{\odot}])}<8.5$ at $z\sim 4$ (at least for GOODS-S; \citealt{duncan2014}), which corresponds to about one-quarter of the sample. We want to stress that this is a homogeneously M$_*$-selected sample, designed to study Ly$\alpha$ emission statistics dependence on galaxy properties. As a consequence, it is by no means representative of the M$_*$ or Ly$\alpha$ LFs at $3<z<4.6$. This must be taken into account when comparing this sample to analogues directly drawn from the galaxy population (i.e. LBGs or narrowband samples). 

\section{Ly$\alpha$ Measurements}
\label{3}
\subsection{Line Detection}
\label{3.1}
For line detection, we use an automated maximum likelihood fitting routine after continuum subtraction. We assume intrinsic Gaussian profiles of the form
\begin{gather}
	\label{gaussian_line}
	f_{rest}(\lambda)=\frac{F}{\sqrt{2\pi \sigma_{\lambda}^{2}}} \mbox{exp} \left[-\frac{(\lambda-\lambda_{0})^{2}}{2\sigma_{\lambda}^{2}}\right]
\end{gather}
where $F$, $\lambda_{0}$, and $\sigma_{\lambda}$ compose the parameter space explored by the maximum likelihood.

In order to account for false positives, we run the line detection routine on the 115 sky fibers. The results are shown in Figure \ref{fig:3}. We detect four lines above 4$\sigma$ and none above 5$\sigma$. Therefore, down to 5$\sigma$, we expect at most two false detections in our $120$ lines, translating into $\lesssim 2\%$ contamination using signal-to-noise S/N$^{*}=5.5$ as our threshold. We also characterize our detection completeness (see Figure \ref{fig:3}). To obtain it, we use $p($S/N$_{i}>$S/N$^{*}|$S/N), with S/N$_{i}$ the measured signal-to-noise ratio and S/N$^{*}$ the imposed detection threshold. We define the simulated signal to noise as S/N$_{sim}$=F$/\sqrt{\sum d\lambda^2 \sigma_k^2}$, with $d\lambda$ the wavelength dispersion in the spectrum and $\sigma_k$ the flux uncertainty for pixel \textit{k}. We find the most accurate representation by summing over an interval of $20$\AA \ centered at 5500 \AA. To recover the actual completeness shown in Figure \ref{fig:3}, we simulate $\sim 10^3$ lines on the 115 sky-spectra sampling fluxes of $10^{-19}-10^{-17}$ ergs s$^{-1}$ cm$^{-2}$, FWHMs between $5-13$ \AA, and wavelengths of 4800-6700 \AA. The fraction for which we measure S/N$_{i}>$S/N$^{*}$ is our completeness. We use S/N$^{*}=5.5$ as our detection threshold, which corresponds to a line-flux sensitivity in the final spectra of $\sim 1\times10^{-17}$ erg s$^{-1}$ cm$^{-2}$ (see Figure \ref{fig:1}).

\subsection{Line Profiles}
\label{3.2}
The radiative transfer and escape of Ly$\alpha$ radiation from galaxies can be highly complicated. As a matter of fact, the resonant nature of this line has led to thorough modeling of its radiative escape (e.g. \citealt{verhamme2006,dijkstra2012}). Such complications imply that the flux profile of a Ly$\alpha$ line is not always well reproduced by the usual Gaussian profile (\citealt{chonis2013,trainor2015}). Hence, for flux measurements, we adopt a more sophisticated model. Similarly to \citet{mclinden2011} and \citet{chonis2013}, we fit double-peaked Gaussian profiles of the form
\begin{gather}
	\label{double-peak}
	f_{rest}(\lambda)= f_{blue}(\lambda) + f_{red}(\lambda)
\end{gather}
where $f_{blue}(\lambda)$ represents the blue emission component and $f_{red}(\lambda)$ represents the red component. We assume each of these components to be asymmetric, i.e., they follow Equation (\ref{gaussian_line}), with $\sigma$ defined as
\begin{gather}
	\label{sigma}
	\sigma_\lambda \equiv \sigma_{b} \mbox{				if	} \lambda < \lambda_0 \\
	\nonumber
	\sigma_\lambda \equiv \sigma_{r} \mbox{				if	} \lambda > \lambda_0 
\end{gather}
Before fitting, we convolve the profiles given by Equations (\ref{double-peak}) and (\ref{sigma}) with the spectral resolution. This allows us to properly characterize the errors in our measurements and methodology. In case that there is some sky contamination (which happens for 28 of the 120 detections), we only fit single-peaked profiles. The resulting fits to six of our emission lines are shown in Figure \ref{fig:1}. 

\subsection{Ly$\alpha$ Equivalent Width and Escape Fraction}
\label{3.3}
There are typically two diagnostics used to characterize the prominence of Ly$\alpha$ emission in galaxies: the equivalent width ($W_{Ly\alpha}$) and the Ly$\alpha$ escape fraction ($f_{esc}$). The rest-frame $W_{Ly\alpha}$ is defined as the fraction between line flux and UV continuum flux in the rest frame of the galaxy. Explicitly,
\begin{equation}
	\label{W_{Ly_alpha}}
	W_{Ly\alpha}=\frac{F}{f_{\lambda}}\frac{1}{(1+z_{Ly\alpha})}
\end{equation}
with $F$ the Ly$\alpha$ flux we measure in the spectra and $f_{\lambda}$ the observed flux at rest frame $1700$\AA \ from 3D-HST rest-frame colors (\citealt{skelton2014}). It must be noted that 3D-HST rest-frame colors come from the best-fit template to the photometry, i.e., they are not direct measurements. Still, this allows us to use, in principle, the same rest-frame wavelength for every object, regardless of its actual redshift. In addition, we can also derive a value of $f_{\lambda}$ even if the rest frame $1700$\AA \ photometry is missing. For reference, we find that the templates to match the photometry at $\sim 1700$\AA \ of every object to a typical agreement of $\sim 90\%$. For the uncertainties on $f_{\lambda}$, however, we use the errors on the photometry. We do not use any $f_{\lambda}$ directly measured from our data, since every galaxy has a continuum fainter or comparable to the 1$\sigma$ errors in the spectra.

On the other hand, $f_{esc}$ is the fraction of the number of Ly$\alpha$ photons that escape the galaxy from the number produced. This diagnostic is typically indirectly recovered using SFRs derived from the Ly$\alpha$ line and intrinsic SFRs. The latter are typically calculated from UV continuum measurements or H$\alpha$ fluxes, subject to extinction correction. In our case, we use the intrinsic (i.e., extinction-corrected) SFRs from FAST. The observed SFRs are derived from the SED, whereas the extinction corrections assume a \citet{calzetti2000} attenuation law with $R_{V}=4.05$. The explicit definition is then (e.g. \citealt{blanc2011})
\begin{gather}
	\nonumber
	f_{esc}(Ly\alpha)= \frac{L(Ly\alpha)_{obs}}{L(Ly\alpha)_{intrinsic}}= \\
	\label{fesc}
	\frac{\mbox{SFR}(Ly\alpha)}{\mbox{SFR}(\mbox{UV})_{corr}}=\frac{4.4\times 10^{-42}L_{Ly\alpha}/8.7}{\mbox{SFR}(\mbox{UV})_{corr}}
\end{gather}
with $L_{Ly\alpha}$ the luminosity associated with the Ly$\alpha$ line and SFR(UV)$_{corr}$ the extinction-corrected SFRs. This assumes a Ly$\alpha$ to H$\alpha$ ratio of 8.7 and the H$\alpha$ SFR calibration for a \citet{chabrier2003} IMF of $4.4\times 10^{-42}$ (\citealt{mathee2016}). In case of non-detections, we use the corresponding S/N$^*=5.5$ line fluxes to derive upper limits on $W_{Ly\alpha}$ and $f_{esc}$. Our calculations do not account for the UV excess associated with binaries, which induce uncertainties on $f_{esc}$ that can be very difficult to account for (\citealt{stanway2016}).

\subsection{AGN contamination}

We do not measure any objects to have more than one emission line with S/N$_{i}>$S/N$^{*}$, dismissing the existence of evident active galactic nuclei (AGNs) in our sample. We do not find any of the 120 Ly$\alpha$-emitting galaxies to have S/N$>1$ emission potentially associated with the C \small{IV}\normalsize $\lambda1549$ line. Moreover, we find no broad-line Ly$\alpha$ emission ($\Delta v>1000$ km s$^{-1}$), consistent with the rarity of Type I AGNs at high redshift (\citealt{dawson2004}), and in strong contrast with the AGN fractions at low redshift ($\sim 0.4$; \citealt{finkelstein2009}). This is consistent with the fact that bright AGNs are found in $\lesssim$ 1\% of $z>3$ LAEs (\citealt{malhotra2003,wang2004,gawiser2007,ouchi2008,zheng2010}). Low-luminosity AGN contamination in LAE samples is more difficult to constrain, but \citet{wang2004} estimate the total AGN fraction to be $<17\%$. We use our $f_{esc}$ measurements to further dismiss the existence of evident AGNs. We perform cross-matching with the NASA/IPAC Extragalactic Database (NED\footnote{The NASA/IPAC Extragalactic Database (NED) is operated by the Jet Propulsion Laboratory, California Institute of Technology, under contract with the National Aeronautics and Space Administration.}) for all objects with $f_{esc}>1$, and we find none of them to be a reported X-ray source.

\section{The Ly$\alpha$ Equivalent Width Distribution}
\label{4}
\subsection{Bayesian Inference}
\label{4.1}
Measurement of $W_{Ly\alpha}$ for a galaxy sample yields the $W_{Ly\alpha}$ distribution. Since $W_{Ly\alpha}$ is directly measured from the data, characterization of the $W_{Ly\alpha}$ distribution can be naively considered straightforward. However, careful consideration of uncertainties and completeness can yield important insights into the underlying information.  Therefore, for proper characterization of uncertainties, significance, and trends in our results, we use Bayesian statistics. In this section, we explain how to recover the $W_{Ly\alpha}$ distribution within this framework, complementary to the one introduced in \citet{treu2012}.

Different probability distribution models can be adopted to reproduce $W_{Ly\alpha}$ distribution measurements. For instance, studies use Gaussian (e.g. \citealt{guaita2010}), exponential (e.g.\ \citealt{jiang2013a,zheng2014}), and log-normal distributions. For our analysis, we define the probability distribution as $p(W_{Ly\alpha}|W_n)$. Let $W_n$ be the parameter space associated with the model. From now on, we describe the Bayesian approach to recover the posterior distribution of $W_n$. By means of this approach, we include the uncertainties in sample size, flux measurements, and photometry in the estimation of the posterior. The description we provide is not limited to this particular work, allowing for further application in similar datasets. We only present here the fundamental equations, as this procedure is already described in detail in \citet{oyarzun2016}. 

Our Bayesian analysis is based on the Ly$\alpha$ line flux $F$ instead of $W_{Ly\alpha}$, as introduced in equation (\ref{W_{Ly_alpha}}). This approach simplifies the equations, since we can assume $F$ and $f_{\lambda}$ to be normally distributed, which cannot be done for $W_{Ly\alpha}$. According to Bayes' theorem, the posterior distribution $p(W_n| \{{\mbox{$F$}}\} )$, i.e., the parameter space probability distribution given our data set $\{ F\}$, is
\begin{gather}
	p(W_n| \{ F\} )= \frac{p(\{ F\}|W_n)p(W_n)}{p(\{ F\} )}
	\label{bayes}
\end{gather}
The likelihood is just the product of the individual likelihood for every galaxy, i.e., $p(\{F\} |W_n)=  \displaystyle \prod p(F_{i} |W_n)$. For a detection, it is given by
\begin{gather}
	p(F_{i}|W_n)  =
	\int_{0}^{\infty} p(F_{i}|F)p(F|W_n)dF
	\label{single}
\end{gather}
where $p(F_{i}|F)$ is the line-flux probability distribution for the corresponding galaxy valued at a flux $F$, which we consider to distribute normally. On the other hand, the term $p(F|W_n)$ is just the probability distribution of $F$ given $W_n$. Using the definition of $W_{Ly\alpha}$ from Equation (\ref{W_{Ly_alpha}}), the term translates to the product distribution $p(F|W_n)=p(W_{Ly\alpha}|W_n) p(f_{\lambda})$. At this point, we include the probability distribution for the continuum, which we also assume to be Gaussian.

The limiting line flux $F^{*}_{i}$ for discerning detections from noise is given by our S/N threshold, i.e., $F^{*}_{i}=$S/N$^{*}\sigma_{i}$. For galaxies with no detections above $F^{*}_{i}$, we adopt the following value for the likelihood:
\begin{gather}
	\nonumber
	p(F_{i}<F^{*}_{i}|W_n)= \\ \int_{0}^{\infty}\left(1-p(F_{i}>F^{*}_{i}|F)\right)p(F|W_n)dF
	\label{nondet}
\end{gather}
with $p(F_{i}>F^{*}_{i}|F)$ the detection completeness at a line flux $F$ (see Section \ref{3.1}).

Using the expressions for detections and non-detections, the posterior distribution takes the final form:
\begin{gather}
	\nonumber
	p(W_n| \{F\})= \\  \frac{p(W_n)}{p(\{F\})}\displaystyle \prod_{D}p(F_{i}|W_n) \displaystyle \prod_{ND}p(F_{i}<F^{*}_{i}|W_n)
	\label{final}
\end{gather}
with $p(W_n)$ the priors of the model parameters and $p(\{F\})$ a normalization constant reflecting the likelihood of the model. The form of Equation (\ref{final}) is general and will be used as a starting point for multiple analysis throughout.

\begin{deluxetable*}{c|ccc|ccc|ccc|ccc}
	\centering
	\tablecaption{Ly$\alpha$ Equivalent Width distribution model comparison}
	\tabletypesize{\footnotesize}
	\tablewidth{0pt}
	\tablehead{
		Likelihood & \multicolumn{3}{c}{Low Mass} & \multicolumn{3}{c}{Medium Mass} & \multicolumn{3}{c}{High Mass} & \multicolumn{3}{c}{Complete Sample} \\
		Output & \colhead{Exp} & \colhead{Gaussian} & \colhead{Log n}& \colhead{Exp} & \colhead{Gaussian} & \colhead{Log n}& \colhead{Exp} & \colhead{Gaussian} & \colhead{Log n}& \colhead{Exp} & \colhead{Gaussian} & \colhead{Log n}}
	\startdata
	Model odds\tablenotemark{a} & 0.93 & 0.06 & 0.01& 0.97 & 0.01 & 0.02& 1 & $10^{-3}$ & $10^{-5}$ & 0.98 & $10^{-6}$ & 0.02\\	
	Peak odds\tablenotemark{b} & 0.25 & 0.02 & 0.73& 0.23 & $10^{-3}$ & 0.77& 0.97 & $10^{-3}$ & 0.03& 0.19 & $10^{-7}$ & 0.81\\
	$A$\tablenotemark{c} & 0.75 & 0.55 & 0.55& 0.55 & 0.4 & 0.5& 0.35 & 0.2 & 0.2& 0.4 & 0.3 & 0.45\\
	$W_0$\tablenotemark{c} & 46 & 74 & 0.7& 26 & 46 & 0.85 & 14 & 26 & 0.75 & 38 & 64 & 1.05\\		
	$\mu$\tablenotemark{c} & ... & ... & 3.85 & ...& ... & 3.05& ... & ... & 3 & ... & ... & 3.05	
	\enddata
	\tablenotetext{a}{Obtained by integrating the likelihood over the whole parameter space.}
	\tablenotetext{b}{Calculated with the maximum of the likelihood.}
	\tablenotetext{c}{Model parameters for the corresponding likelihood maximum. Note that these values are different from the ones in \citet{oyarzun2016}, since here we are just working with the likelihood.}	
	\label{table:1}
\end{deluxetable*}

\subsection{Model Comparison}
\label{4.2}
As we introduced in the previous section, multiple probability distributions can be adopted for the representation of the $W_{Ly\alpha}$ distribution. In order to perform model selection, several elements are taken into account, such as model complexity and number of parameters. In this section, we describe our methodology to perform such a selection from a quantitative standpoint. By means of a Bayesian approach, we recover probability ratios for the different models, providing insight into how we perform the selection given our measurements. The analysis presented here is a quantitative implementation of Occam's Razor and is not unique to our dataset, i.e., it can be applied to any dataset modeling. 

Every model we discuss here is composed of a scaled probability distribution and a Dirac delta, as proposed in \citet{treu2012}. If we define the standard probability distributions as $p_0(W_{Ly\alpha}|W_n)$, the modified counterparts we consider are given by
\begin{gather}
	p(W_{Ly\alpha}|W_n)=A\times H(W_{Ly\alpha}) \times p_0(W_{Ly\alpha}|W_n)\\
	\nonumber
	+(1-A)\times \delta(W_{Ly\alpha})
\end{gather}

where the first term is the scaled probability distribution. It is multiplied by the Heaviside $H(W_{Ly\alpha})$ to ensure it only represents positive $W_{Ly\alpha}$ values. Hence, this scaled term integrates $A$, i.e., $A$ can only adopt values between 0 and 1. Note that this term is different from the fraction of detections in our sample, as we consider upper limits for our non-detections. The second term groups the fraction of galaxies that do not emit in Ly$\alpha$ (i.e., no line and/or absorption). As our data is restricted to emission lines, we represent this term using the Dirac delta. 

We explore here exponential-, Gaussian-, and log-normal- based distributions. The first two are two-parameter models, while the log-normal is three-parameter dependent. Hence, we generalize our parameter space as $W_n=(A, W_{0}, \mu)$. Then, the expressions for our exponential, Gaussian, and log-normal models are, respectively,

\begin{gather}
	\nonumber
	p(W_{Ly\alpha}|\mbox{exp})=\frac{A}{W_{0}}e^{-W_{Ly\alpha}/W_{0}}H(W_{Ly\alpha})\\
		\label{exponential}
	+(1-A)\delta(W_{Ly\alpha})
\end{gather}
\begin{gather}
	\label{gaussian}	
	\nonumber
	p(W_{Ly\alpha}|\mbox{Gauss})=\frac{2A}{\sqrt{2\pi W_0^2}}e^{-W_{Ly\alpha}^{2}/2W_0^{2}}H(W_{Ly\alpha})\\
	+(1-A)\delta (W_{Ly\alpha})
\end{gather}
\begin{gather}
	\label{lognormal}	
	\nonumber
	p(W_{Ly\alpha}|\mbox{log n})=\frac{A}{\sqrt{2\pi W_0^2W_{Ly\alpha}^2}}e^{-(ln(W_{Ly\alpha})-\mu)^2/2W_0^2}\\
	+(1-A)\delta(W_{Ly\alpha})
\end{gather}
We now describe our approach for comparing the three models. For a set of measurements, Bayes' theorem gives the probability of model $M_i$ in the model space $\{M\}$ 
\begin{gather}
	p(M_i | \{ F\})=  \frac{p(\{ F\}|M_i)p(M_i)}{p(\{ F\})}
\end{gather}
with $p(M_i)$ the prior for model $M_i$ in the set $\{M\}$, which we assume to be equal for the three models. Once again, $p(\{ F\})$ is a normalization constant. Therefore, the probability for model $M_i$ is proportional to the likelihood of the model, i.e.,
\begin{gather}
	p(M_i | \{ F\}) \propto p(\{ F\}|M_i) = \int_{n}p(W_n)p(\{F\}|W_n)dW_n
	\label{model_likelihood}
\end{gather}
The absolute probability of each model $M_i$ within the set $\{M\}$ is obtained by imposing that the models explored cover all possible choices, i.e., $\sum p(M_i | \{ F\})=1$. Hence, the probability of $M_i$ given our dataset is
\begin{gather}
	p(M_i | \{ F\}) = \frac{ \int_{n}p(W_n)p(\{F\}|W_n)dW_n}{\sum p(M_i | \{ F\})}
	\label{model_theorem}
\end{gather}
For an analytically correct model comparison, analysis of Equation (\ref{model_theorem}) is required. Still, the term in Equation (\ref{model_likelihood}) is strongly dependent on the priors assumed for the parameter space $W_n$ of every model. Therefore, different prior selections can have significant effects on the odds for each model.  As a workaround, we rewrite the model probabilities as
\begin{gather}
	p(M_i | \{ F\}) = \frac{ \int_{n}p(\{F\}|W_n)dW_n}{\sum p(M_i | \{ F\})}
	\label{odds}
\end{gather}
Effectively, this simplification conveniently limits our analysis to a pure likelihood comparison, i.e., we adopt constant, uninformative priors. This is equivalent to assuming ignorance in linear scales for every parameter. Since our distributions are smooth and single peaked, we are confident in this assumption. The dimensions of our parameter spaces do not go beyond three, and the uncertainties in our parameters are of the order of the most probable values, providing further assurance to our assumptions.

\begin{figure*}
	\centering
	\includegraphics[width=7in]{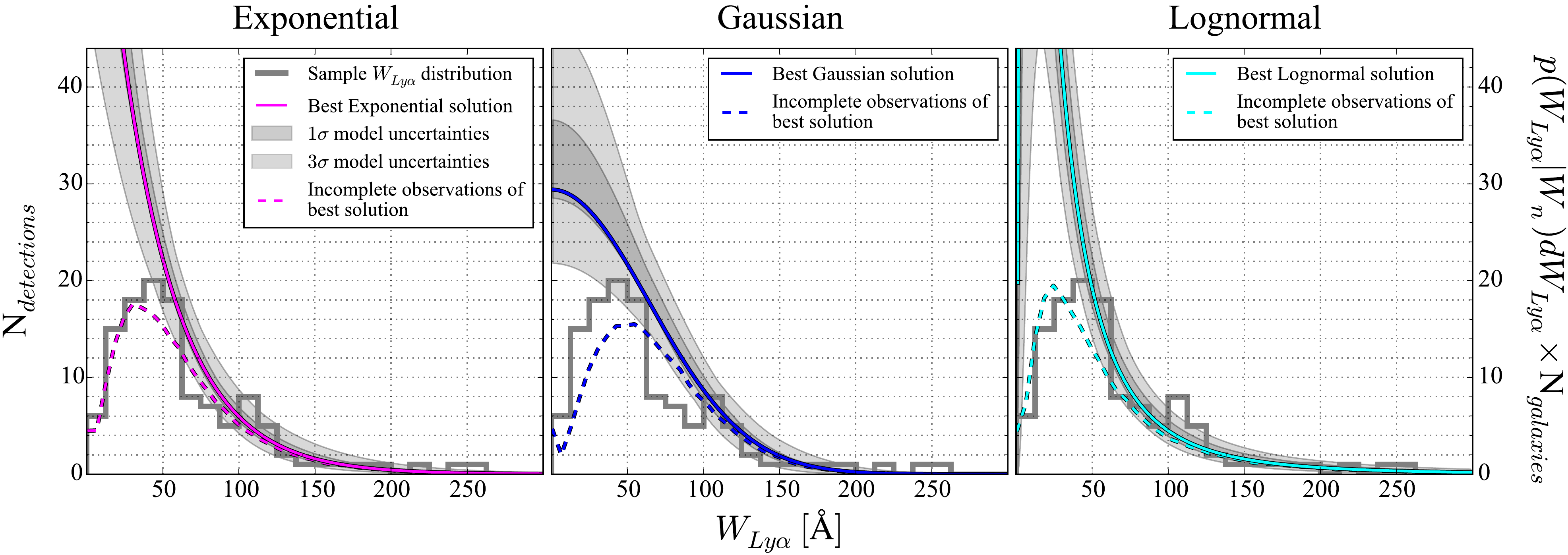}
	\caption{Shown in histograms are the observed rest-frame $W_{Ly\alpha}$ distribution for the whole sample. These histograms account for the 120 measured $W_{Ly\alpha}$ with S/N$\geqslant$5.5. From left to right, we overplot the exponential, Gaussian, and log-normal model constraints, respectively. The solid colored lines correspond to the peak of the likelihood, i.e., the most probable solutions. The dashed lines represent these solutions corrected for our completeness. The shaded regions show the $1\sigma$ and $3\sigma$ confidence levels yielded by the likelihood. Our Bayesian model comparison prefers the exponential model, since it reproduces our $W_{Ly\alpha}$ measurements fairly well despite having only two free parameters.}	
	\label{fig:4}	
\end{figure*}

In \citet{oyarzun2016}, we show that a galaxy sample with a broad M$_*$ range yields a composite $W_{Ly\alpha}$ distribution. For the rest of this section, we divide our sample into three M$_*$ bins, as shown in Figure \ref{fig:2}. We use the complete sample and these three subsamples to contrast the models. The outcomes are presented in Table \ref{table:1} and Figure \ref{fig:4}. Table \ref{table:1} gives evidence that the best likelihood is obtained with the log-normal model for three of the four distributions. This can be verified in our distribution simulations for the complete sample in Figure \ref{fig:4}, especially toward the high $W_{Ly\alpha}$ tail. Still, when integrating the likelihoods, the lognormal distribution is the least probable. This is a consequence of the extra parameter needed by the model, which penalizes the likelihood when integrating over the parameter space. Then, according to our analysis, the preferred model is the exponential. While it models the distribution better than the Gaussian, it also reproduces our $W_{Ly\alpha}$ measurements fairly well, despite depending on only two parameters. In addition, the uncertainties in Figure \ref{fig:4}, especially for low $W_{Ly\alpha}$, confirm this model is the most adequate to reproduce our measurements. We remark that the procedure for model selection described here considers the lower $W_{Ly\alpha}$ end of the distribution, which includes our completeness and non-detections. Nonetheless, further $W_{Ly\alpha}$ distribution analysis in this paper is mostly focused on the higher $W_{Ly\alpha}$ end and is not strongly dependent on model preference. 

From now on, we perform our $W_{Ly\alpha}$ distribution analysis using the exponential model of Equation (\ref{exponential}). We stress that this expression is dependent on the parameters $A$ and $W_0$, with the first being the fraction of galaxies showing emission and the second the e-folding scale of the distribution. We advise caution when using $A$ as a proxy for the fraction of line emitters in the parent population, since it is tied to the adequacy of an exponential profile.

\section{Ly$\alpha$ Emission Dependence on Galaxy Properties}
\label{5}
\subsection{Stellar Mass}
\label{5.1}
Evidence suggests Ly$\alpha$ emission is strongly dependent on the M$_*$ of galaxies. Galaxies with higher M$_*$ have been forming stars for longer, leading to greater ISM dust that presumably forms in supernovae and AGB stars (\citealt{silva1998}). A greater dust content leads to more Ly$\alpha$ photon absorption, decreasing $W_{Ly\alpha}$. This effect has already been observed, at least for high $W_{Ly\alpha}$, in \citet{blanc2011} and \citet{hagen2014}. Similarly, the bulk of M$_*$ is dominated by older stars, which do not contribute significantly to the Ly$\alpha$ photon budget of galaxies. As a matter of fact, Ly$\alpha$ emission decreases steadily with the age of stellar populations, as seen in \citet{charlot1993} and \citet{schaerer2003}. Ly$\alpha$ radiative transfer is also severely affected by the neutral gas structure and kinematics of the ISM and circumgalactic medium (\citealt{verhamme2006}). Since more massive star-forming galaxies are bound to have higher gas mass (e.g. \citealt{keres2005,finlator2007}), Ly$\alpha$ photons should be subject to more resonant scattering, therefore decreasing their $W_{Ly\alpha}$. The trends we find in \citet{oyarzun2016} confirm this qualitative scheme at $z\sim 4$. In this section, we perform a more detailed and robust characterization of Ly$\alpha$ emission dependence on M$_*$.

\begin{figure*}
	\centering
	\includegraphics[width=7in]{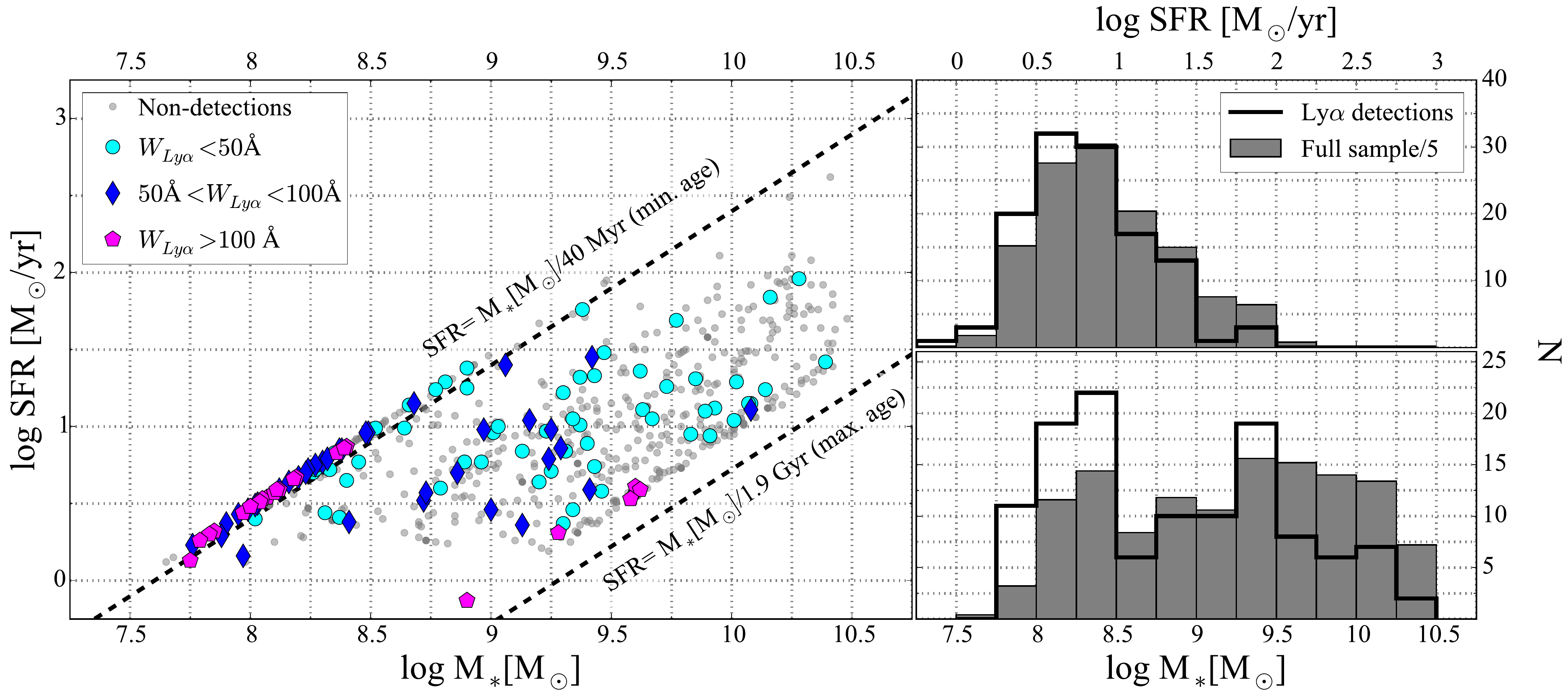}
	\caption{Detail of the M$_*$ and SFR of our galaxies. Left: M$_*$-SFR plane location of our objects. Non-detections are shown in gray. Low, medium, and high $W_{Ly\alpha}$ detections are plotted as cyan circles, blue diamonds, and magenta pentagons, respectively. The values of M$_*$ and SFR we show come from SED fitting of 3D-HST photometry using FAST under the assumption of constant SFHs. The dashed lines indicate the minimum and maximum M$_*$ that galaxies can reach assuming that they have been forming stars at a constant rate. Right: M$_*$ (top) and SFR (bottom) histograms of the 625 galaxies in our sample (grey) and 120 detections (black). For better comparison, we divide the histogram values for the full sample by five. Note that inferences on the fraction of detections must take into account that incompleteness is dependent on both M$_*$ and SFR. In this paper, we use the exponential W$_{Ly\alpha}$ distribution normalization ($A$; Equations \ref{mass1} and \ref{mass2}) as a proxy for the dependence of the Ly$\alpha$ fraction on M$_*$. Our measurements are more than $2\sigma$ consistent with a decrease in the fraction going to higher M$_*$ (see Figure \ref{fig:7}).}	
	\label{fig:5}
\end{figure*}

\begin{figure}
	\centering
	\includegraphics[width=3.4in]{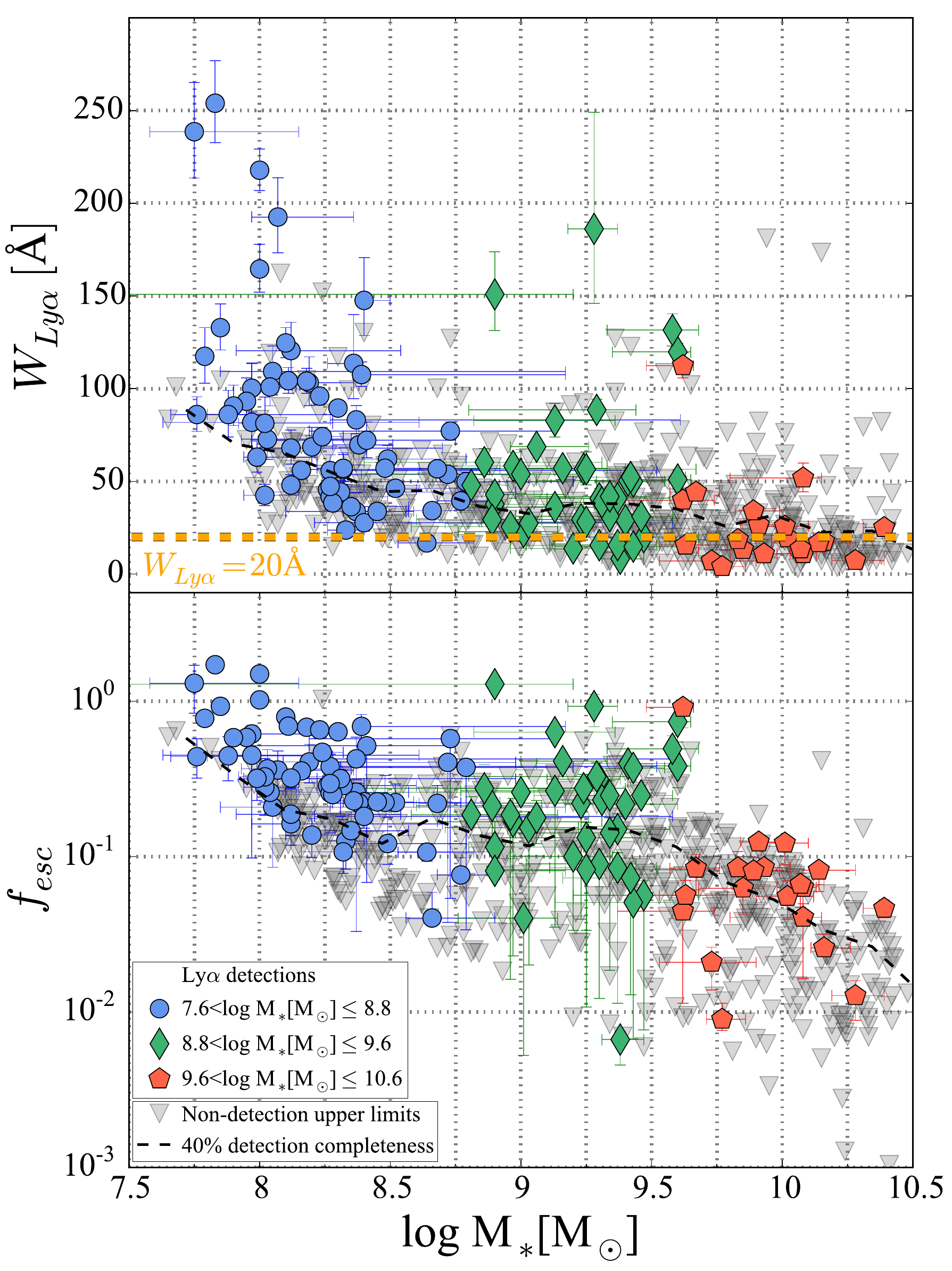}
	\caption[Ly$\alpha$ Equivalent Width and Escape Fraction dependence on Stellar Mass]{Top: plot of our $W_{Ly\alpha}$ as a function of M$_*$. Bottom: Plot of our Ly$\alpha$ escape fractions as a function of M$_*$. The detections corresponding to the low-, medium-, and high-mass subsamples are shown as blue circles, green diamonds, and red pentagons, respectively. We observe both Ly$\alpha$ diagnostics,  $W_{Ly\alpha}$ and $f_{esc}$, to anti-correlate with M$_*$. The upper limits associated with non-detections are plotted as gray triangles. We use these non-detections to give a rough estimate of our 40\% completeness (dashed). Note how our completeness is not independent of M$_*$, which has to be considered when making inferences on the M$_*$ distribution and characteristic $W_{Ly\alpha}$ ($f_{esc}$) of LAEs. We also show in yellow the typical rest-frame selection threshold of $W_{Ly\alpha}>20$\AA \ imposed by narrowband surveys (e.g. \citealt{gronwall2007,guaita2010,adams2011}).}	
	\label{fig:6}
\end{figure}

Our sample is especially designed to study the dependence of Ly$\alpha$ emission in M$_*$. As shown in Figure \ref{fig:2}, our objects are selected in redshift and M$_*$, homogeneously covering the range $7.6<log(\mbox{M}_*[\mbox{M}_{\odot}])<10.6$. For further clarity, we plot in Figure \ref{fig:5} the M$_*$ and SFRs of our complete sample and detections. Comparison of the M$_*$ histograms slightly hints at an anti-correlation between the LAE fraction and M$_*$, at least down to our detection limit. However, since our completeness depends on M$_*$, a more thorough analysis is required (see below). The existence of any Ly$\alpha$ emission dependence on M$_*$ becomes clearer in Figure \ref{fig:6}, where we plot $W_{Ly\alpha}$ and $f_{esc}$ as a function of M$_*$. We also plot in Figure \ref{fig:6} our upper limits for non-detections. The regions sampled by our non-detections reveal how our completeness is not independent of M$_*$. Our detections are flux limited, so we achieve lower $W_{Ly\alpha}$ for galaxies with brighter UV continuum. Therefore, even though we observe lower M$_*$ galaxies to have higher $W_{Ly\alpha}$ and $f_{esc}$, any qualitative conclusions we can draw involving the fraction of LAEs as a function of M$_*$ are affected by our completeness. Still, there is a clear upper envelope to the distribution of galaxies in this plot, where we are not affected by incompleteness. For M$_*<10^{8.5}$, there is a clear anti-correlation between $W_{Ly\alpha}$ and M$_*$. For more massive galaxies, however, the trend is mostly flat, except for the presence of a few interlopers. Still, our qualitative result is that both Ly$\alpha$ emission diagnostics show an anti-correlation with M$_*$. For the rest of this section, we focus on how to thoroughly quantify the effect of M$_*$ on $W_{Ly\alpha}$.

\begin{figure*}
	\centering
	\includegraphics[width=7in]{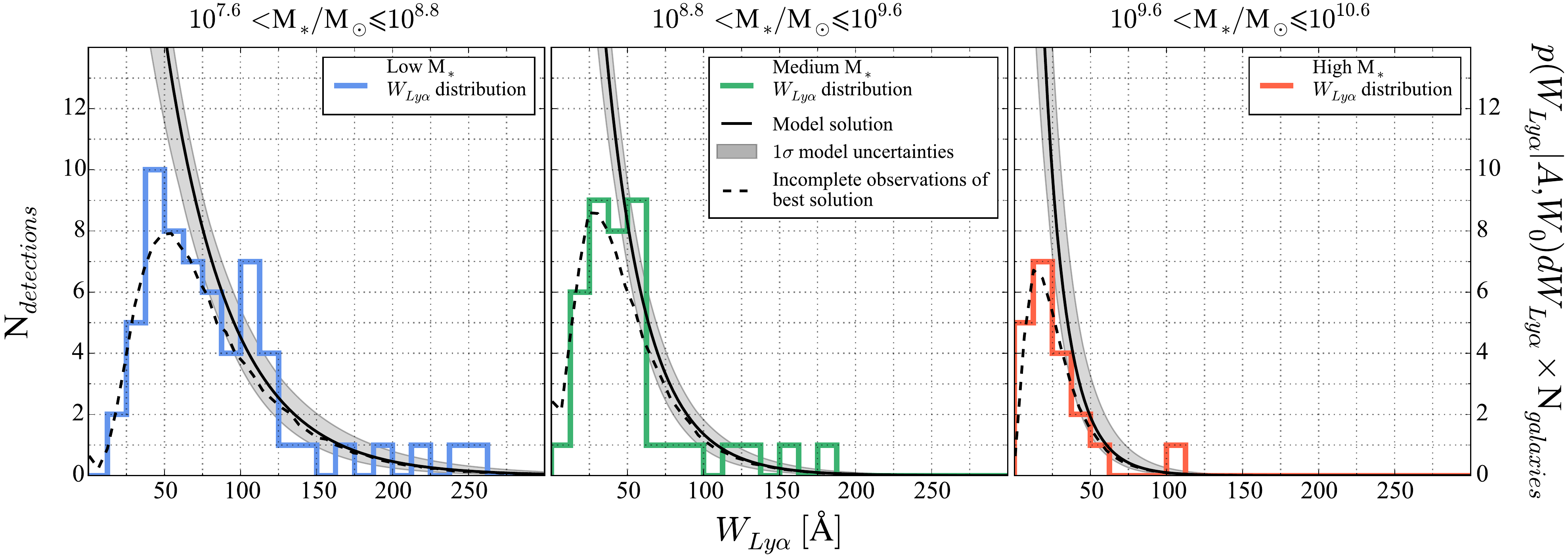}
	\caption{Rest-frame $W_{Ly\alpha}$ distributions of the low-, medium-, and high-mass bins, from left to right. We use Monte Carlo simulations to characterize the exponential $W_{Ly\alpha}$ distribution dependence on M$_*$ following Equations (\ref{mass1}) and (\ref{mass2}). We can then simulate $W_{Ly\alpha}$ distributions for every subsample and obtain 1$\sigma$ constraints (shaded contours). Keep in mind that these constraints are given by collections of curves resulting from our Monte Carlo simulations. Our best results correspond to the median distribution we expect by considering every object in the bin. We also plot as dotted lines the completeness-corrected counterparts of the median distributions. In contrast to \citet{oyarzun2016}, where the modeling is based on the characteristic M$_*$ of each bin, the results here take into account the M$_*$ of every galaxy. As we show in Equations (\ref{mass12}) and (\ref{mass22}), we find both parameters, the fraction of Ly$\alpha$-emitting galaxies and characteristic $W_{Ly\alpha}$ scale, to anti-correlate with M$_*$ with significances higher than 2$\sigma$ and 6$\sigma$, respectively.}	
	\label{fig:7}
\end{figure*}

The overall dependence of Ly$\alpha$ emission on M$_*$ implies that $W_{Ly\alpha}$ distributions in the literature (e.g., \citealt{gronwall2007,zheng2014}) are influenced by the M$_*$ distribution of the sample. In comparison to deeper M$_{UV}$ surveys, shallower samples are bound to observe lower $W_{Ly\alpha}$. This can lead to incorrect contrast of surveys and misinterpretation of trends. To verify these claims, we divide our sample into three M$_*$ bins (see Figure \ref{fig:2}) and plot the resulting $W_{Ly\alpha}$ distributions in Figure \ref{fig:7}. As expected, there is an apparent anti-correlation between M$_*$ and both, the tail of the $W_{Ly\alpha}$ distribution and the normalization. We perform our first quantitative characterization of the $W_{Ly\alpha}$ distribution dependence on M$_*$ in \citet{oyarzun2016}. In that study, we divide our sample into three M$_*$ bins and obtain the posterior distribution for the exponential parameters separately. This procedure allowed us to fit a linear relation to the final parameters, recovering $A$(M$_*$) and $W_0$(M$_*$) using expressions of the form
\begin{gather}
	\label{mass1}
	A(\mbox{M}_*) = A_{\mbox{\scriptsize M}_*}\log(\mbox{M}_*[\mbox{M}_{\odot}])+A_C \\
	W_{0} (\mbox{M}_*) = W_{\mbox{\scriptsize M}_*}\log(\mbox{M}_*[\mbox{M}_{\odot}]) + W_C
	\label{mass2}
\end{gather}
In this section, we recover these linear relations directly from the complete sample, i.e., we recover the posterior distribution for the four-parameter space composed of the linear coefficients in Equations (\ref{mass1}) and (\ref{mass2}). This more robust methodology does not rely on binning, while also allowing us to constrain the errors on the coefficients directly from the model and measurements. We once again start from Equation (\ref{final}). As mentioned, our parameter space is now $W_n=(A_{\mbox{\scriptsize M}_*}, A_C, W_{\mbox{\scriptsize M}_*}, W_C)$. As these linear coefficients represent the exponential parameters of Equation (\ref{exponential}), deriving non-informative priors is highly complicated. Therefore, in order to determine our priors, we only consider linear scales ignorance. Therefore, the posterior translates to
\begin{gather}
	p(A_{\mbox{\scriptsize M}_*}, A_C, W_{\mbox{\scriptsize M}_*}, W_C| \{F\})= C\times \\
	\nonumber
	\displaystyle \prod_{D}p(F_{i}|A(\mbox{M}_*), W_{0}(\mbox{M}_*)) \displaystyle \prod_{ND}p(F_{i}<F^{*}_{i}|A(\mbox{M}_*),W_{0}(\mbox{M}_*))
\end{gather}
with $C$ a normalization constant. 

We use MCMC simulations to characterize this four-parameter posterior. Its maximum gives the best solution for $A(\mbox{M}_*)$ and $W_0(\mbox{M}_*)$, while the collapsed posteriors yield the uncertainties on the parameters. We can then write Equations (\ref{mass1}) and (\ref{mass2}) as
\begin{gather}
	\label{mass12}
	A(\mbox{M}_*) = -0.28_{-.02}^{+.12}\log{(\mbox{M}_*[\mbox{M}_{\odot}])}+ 3.1^{+0.15}_{-1.1} \\
	W_{0} (\mbox{M}_*) = -17.7_{-2.0}^{+2.6} \log{(\mbox{M}_*[\mbox{M}_{\odot}])} + 190^{+20}_{-25}
	\label{mass22}
\end{gather}

In the framework of an exponential profile, these relations we recover yield the $W_{Ly\alpha}$ probability distribution for an object with known M$_*$. Hence, we can simulate the expected $W_{Ly\alpha}$ distribution for each of the three M$_*$ subsamples and compare with our direct measurements. The results are presented in Figure \ref{fig:7}. Our constraints are consistent with the observed $W_{Ly\alpha}$ distributions.

Our results regarding the dependence of Ly$\alpha$ on M$_*$ are conclusive. In the range $10^{8}-10^{10.5}\mbox{M}_{\odot}$, the $W_{Ly\alpha}$ probability distribution extends to higher $W_{Ly\alpha}$ for lower M$_*$ galaxies. In other words, more massive galaxies tend to have lower $W_{Ly\alpha}$. A similar trend is observed for $f_{esc}$, further highlighting the role of dust and gas mass in the escape of Ly$\alpha$ photons. At $z\sim 2.2$, \citet{mathee2016} also observe the $f_{esc}$ anti-correlation from Figure \ref{fig:6}, although only when stacking galaxies. Their massive objects showing high $f_{esc}$, which they associate with dusty gas outflows, seem to lie below the M$_*$-SFR sequence at $z\sim 2$. These inferences, combined with the significantly higher $f_{esc}$ they measure for larger apertures, make sense in a Ly$\alpha$ diffuse halo scheme, which we do not observe due to our aperture size. Studies on the dependence of $f_{esc}$ on M$_*$ also explore $1.9<z<3.6$ (\citealt{hagen2014}). Their $W_{Ly\alpha}$-selected LAEs follow a trend similar to the one we find at $z\sim 4$. In summary, the evidence for an anti-correlation between $W_{Ly\alpha}$ (or $f_{esc}$) and M$_*$ is significant, but the scatter seems to depend on measurement methodology and sample selection.

Inferences on the M$_*$ distribution of LAEs are not as evident. \citet{hagen2014} do not find their $1.9<z<3.6$ Ly$\alpha$ luminosity-selected LAE number distribution to depend on M$_*$. Their results agree with the $z\sim 3.1$ narrowband-selected survey of \citet{mclinden2014}. However, we show in Figure \ref{fig:6} that spectroscopic completeness is not independent of M$_*$. Therefore, most LAEs survey follow-ups could have higher incompleteness toward lower M$_*$. Since our Bayesian analysis takes into account our completeness for every object, we can test the significance of this claim. The coefficient $A_{\mbox{\scriptsize M}_*}$ in Equation (\ref{mass1}) represents the exponential fraction of LAE dependence on M$_*$. As evidenced by Equation (\ref{mass12}), our measurements are more than $2\sigma$ consistent with a decrease in the fraction going to higher M$_*$. In the scheme of an exponential model, this translates to an LAE distribution dominated by lower M$_*$ galaxies. This result complements the much more significant anti-correlation between $W_{0}$ and M$_*$ we find (see Equation \ref{mass22}).


\begin{figure}
	\centering
	\includegraphics[width=3.4in]{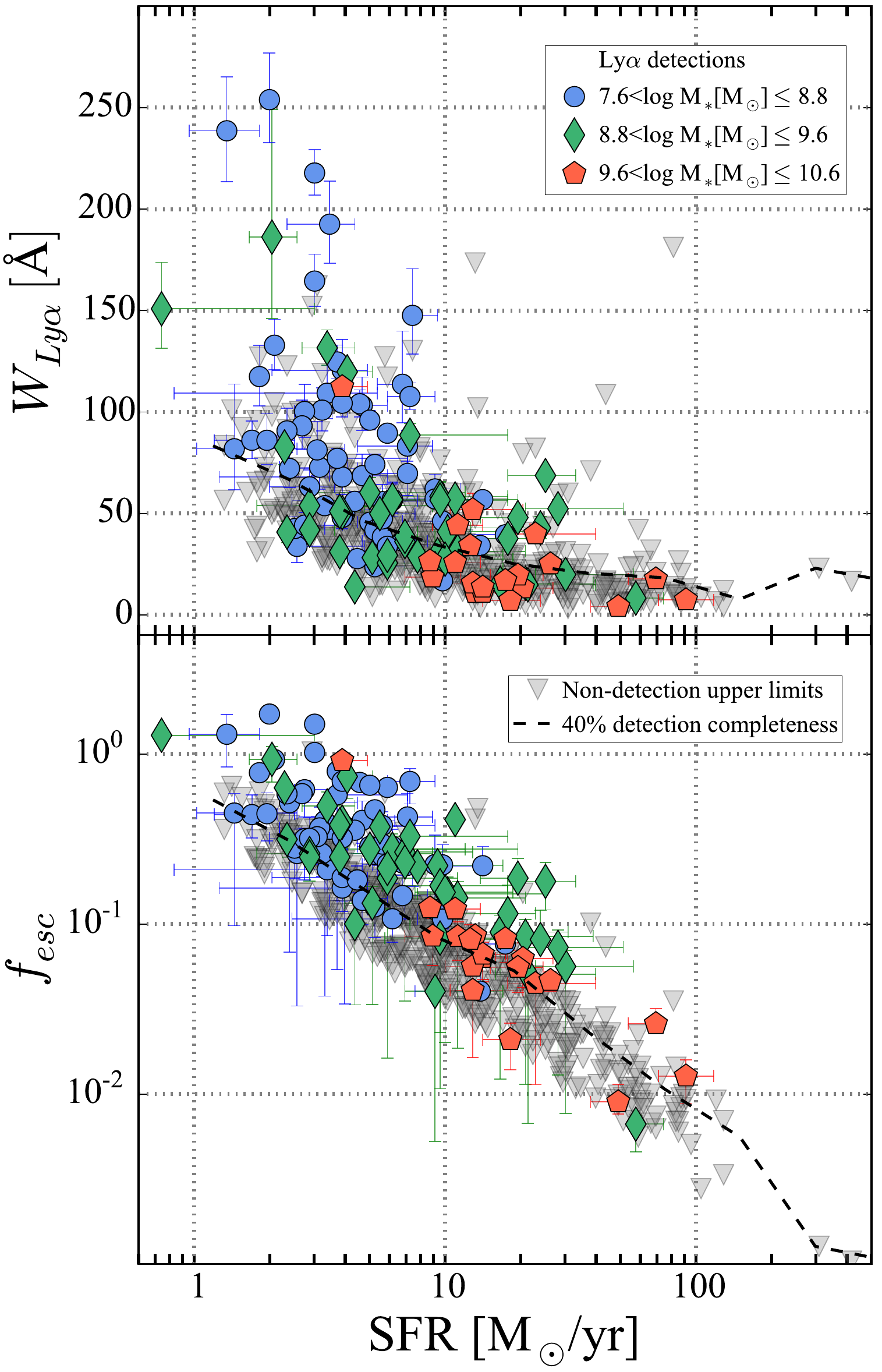}
	\caption{Plot of our $W_{Ly\alpha}$ (top panel) and $f_{esc}$ (bottom panel) as a function of intrinsic SFR. Detections corresponding to our low-, medium-, and high-mass bins are shown as the blue circles, green diamonds, and red pentagons, respectively. We find both measurements to anti-correlate with the SFR we measure from SED fitting. For the plot in the bottom panel, we use the same value for SFRs when determining the escape fractions and galaxy SFRs. For reference, we show our non-detection upper limits as gray triangles. We use these non-detections to give a rough estimate of our 40\% completeness (dashed), confirming how strongly our sensitivity depends on SFR.}	
	\label{fig:9}
\end{figure}
\subsection{SFR}
\label{5.2}
High-redshift galaxies have been observed to follow a correlation between SFR and M$_*$, known as the star-forming main sequence (e.g., \citealt{keres2005,finlator2007,stark2009,gonzalez2011,whitaker2012}; see our Figure \ref{fig:5}). In terms of the underlying physics, more massive objects dominate gas accretion in their neighborhood, feeding and triggering star-formation. Such gas infall seems to dominate over galaxy growth at high redshift (\citealt{keres2005,finlator2007}). This scheme implies that more massive objects form stars at higher rates, at least down to our observational limitations and modeling of high-redshift ISM. Given our results on M$_*$ from the previous section, we expect similar trends between $W_{Ly\alpha}$ ($f_{esc}$) and SFR (Figure \ref{fig:5}). Even more, star-forming galaxies have a higher neutral gas mass, which can hamper the escape of Ly$\alpha$ photons from galaxies (\citealt{verhamme2006}). In fact, it has also been suggested that photoelectric absorption rules Ly$\alpha$ depletion, even over dust attenuation (\citealt{reddy2016}). In this section, we explore any Ly$\alpha$ dependence on SFR within our dataset. We remark that our SFRs come from SED fitting of 3D-HST photometry using FAST (see Section \ref{2.3}), i.e., they have typical associated timescales of 100 Myr (\citealt{kennicutt1998}). We stress that our derived SFRs differ from 3D-HST SFRs, since our calculation assumes cSFHs instead of exponentially declining SFHs (\citealt{skelton2014}). 

We show the $W_{Ly\alpha}$ and $f_{esc}$ dependence on SFR in Figures \ref{fig:5} and \ref{fig:9}. In the latter, we include upper limits for our non-detections to give an insight into how our incompleteness depends on SFR. A clear anti-correlation between $W_{Ly\alpha}$ ($f_{esc}$) and SFR is observed. These results come as no surprise, as they have been previously reported. Most studies of $W_{Ly\alpha}$ dependence on SFR involve uncorrected SFRs (\citealt{pettini2002,shapley2003,yamada2005,gronwall2007,tapken2007,ouchi2008}; all compiled in \citealt{verhamme2008}). Even without dust correction, the anti-correlation is still present in these studies (refer to Figure 19 in \citealt{verhamme2008} and Section \ref{5.4} of this work). Based on a $z\sim 2$ H$\alpha$ emitters sample, \citet{mathee2016} also observe a clear anti-correlation between Ly$\alpha$ $f_{esc}$ and SFR. Interestingly, they do not only observe such a trend in their individual objects, but likewise on their stacks when using different apertures (galaxy diameters of 12 and 24 kpc). As their dataset includes H$\alpha$ fluxes, they can recover SFRs and $f_{esc}$ using H$\alpha$ luminosities. The fact that they observe similar trends with such a different sample suggests that the anti-correlation between $W_{Ly\alpha}$ ($f_{esc}$) and SFR is not only independent of redshift, but also observational constraints like aperture and methodology for recovering SFRs. Their comparison, however, is restricted to SFRs higher than $\sim 5\mbox{M}_{\odot}$yr$^{-1}$. Most of our low-mass objects have SFRs lower than $\sim 5\mbox{M}_{\odot}$yr$^{-1}$, but they seem to follow the same regime as the rest of our sample. Even though $f_{esc}$ uncertainties and incompleteness increase toward lower-SFR, UV-fainter galaxies (Figure \ref{fig:9}), our results suggest that $f_{esc}$ reaches values of 100\% toward SFR$\sim 1-3\mbox{M}_{\odot}$yr$^{-1}$. These numbers are consistent with the analysis by \citet{atek2014}. They compare their $z<0.5$ SFR(Ly$\alpha$)/SFR(UV) measurements with the literature at $z>2$ (\citealt{taniguchi2005,gronwall2007,guaita2010, curtis-lake2012,jiang2013a}). Our finding that $f_{esc}$ reaches values of 100\% toward SFR$\sim 1-3\mbox{M}_{\odot}$/yr overestimates SFR(Ly$\alpha$)/SFR(UV) for $z<0.5$, but is consistent with higher redshift. As pointed by \citet{atek2014}, it seems that $f_{esc}$ at fixed SFR(UV) increases with redshift. Then again, it must be kept in mind that these literature results consider uncorrected SFRs. \\
\medskip


\subsection{UV Luminosity}
\label{5.3}
In this section, we analyze the M$_{UV}$ distribution of our sample, while also exploring any correlations between $W_{Ly\alpha}$ and UV luminosities. It must be noted, though, that our sample is not representative of the galaxy population at $3<z<4$. First, our galaxies are homogeneously distributed in M$_*$, i.e., they are not a random sample from $3<z<4$ CANDELS objects. Second, we are affected by CANDELS completeness, which decreases toward lower M$_*$ galaxies. Since more massive galaxies tend to have higher UV luminosities (\citealt{stark2009, gonzalez2014}), our sample has a higher contribution of bright M$_{UV}$ galaxies than a population-representative subsample.

\begin{figure}
	\centering
	\includegraphics[width=3.4in]{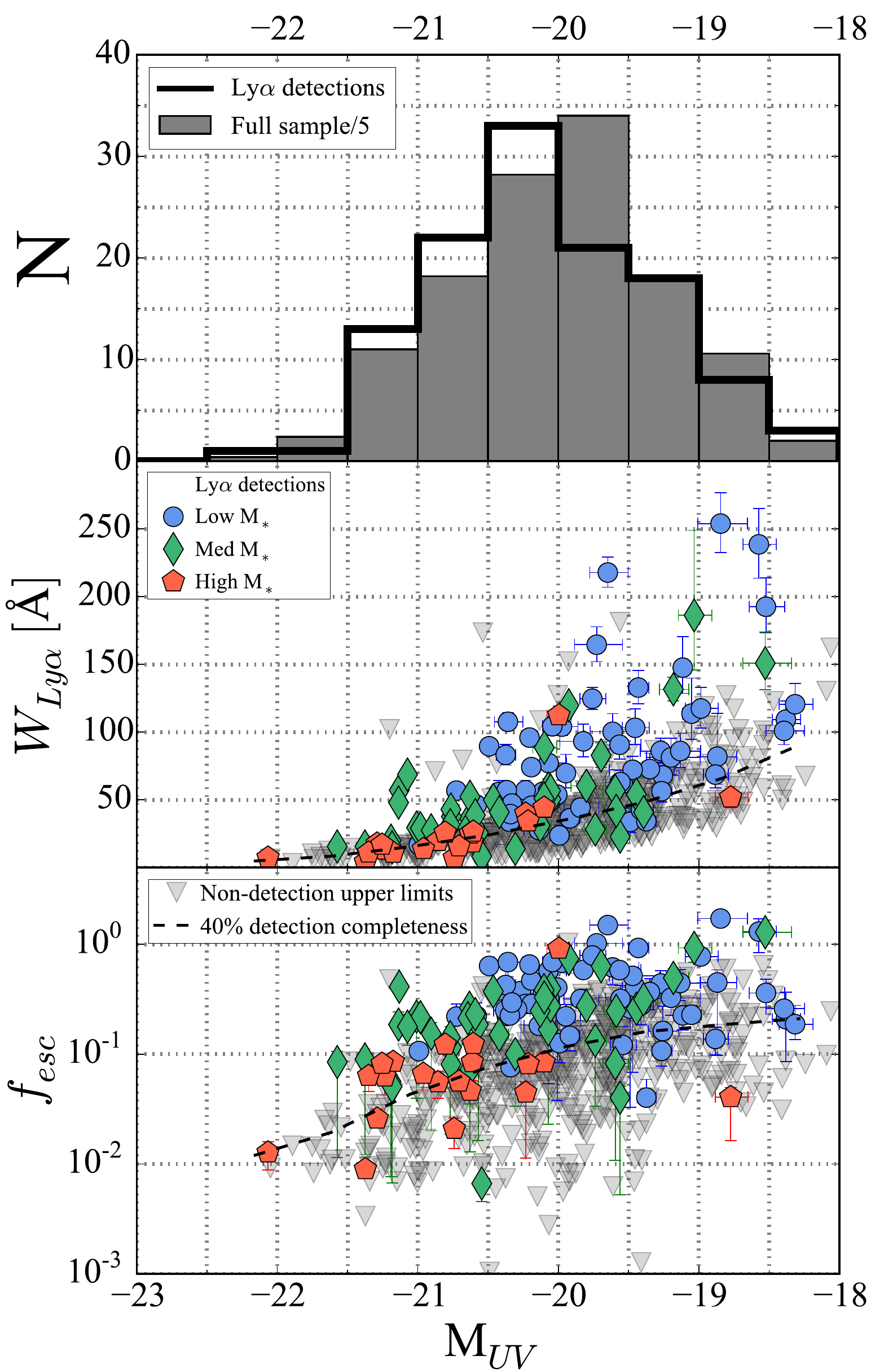}
	\caption{Top: histograms of sample (gray) and detections (black) as a function of UV absolute magnitude. We use the CANDELS $i_{775}$ band for the derivation of M$_{UV}$. For better comparison, we divide the histogram values for the full sample by five. The photometric measurement on this particular band is available for 621 of the 625 galaxies. Center: dependence of $W_{Ly\alpha}$ on UV absolute magnitude. Bottom: dependence of $f_{esc}$ on M$_{UV}$. Blue circles, green diamonds, and red pentagons correspond to low-, medium-, and high-mass subsample detections, respectively. Non-detection upper limits are shown as gray triangles, and we use them to give a rough estimate of our 40\% completeness (dashed). Our results suggest that $W_{Ly\alpha}$ and $f_{esc}$ anti-correlate with UV luminosity. These trends are consistent with studies performed on LBG samples (\citealt{stark2010,schaerer2011}) and narrowband surveys (\citealt{ouchi2008}).}	
	\label{fig:11}
\end{figure}

In order to determine M$_{UV}$ for our objects, we use the CANDELS $i_{775}$ band. We show in Figure \ref{fig:11} the corresponding distribution of the complete sample and detections. We also present the dependence of $W_{Ly\alpha}$ on M$_{UV}$ in this figure. As expected from the SFR-$W_{Ly\alpha}$ anti-correlations we recover in Section \ref{5.2}, a similar trend is observed for UV luminosities. This anti-correlation comes as no surprise, since brighter UV galaxies tend to have higher M$_*$ at the cosmic time of our sample (\citealt{stark2009,gonzalez2014}). Galaxies brighter in the UV have been subject to more intensive star-formation events in 100 Myr timescales. Typically, higher neutral gas, turbulence, and bulk gas motions are associated with higher SFRs, boosting the scatter of Ly$\alpha$ photons. In combination, higher dust extinction, older age of stellar populations, and greater neutral gas mass in UV brighter galaxies seem to rule Ly$\alpha$ statistics dependence on SFR. 

Most analyses of Ly$\alpha$ emission dependence on M$_{UV}$ have been performed using uncorrected SFRs. The literature compilation shown in \citet{verhamme2008} reveals how observed UV SFRs anti-correlate with $W_{Ly\alpha}$ from z$\sim2.7$ to 5.7 (\citealt{pettini2002,shapley2003,yamada2005,gronwall2007,tapken2007,ouchi2008}). Analyses explicitly M$_{UV}$ have also been performed in the surveys of \citet{shimasaku2006}, \citet{ouchi2008}, \citet{vanzella2009}, \citet{balestra2010}, \citet{stark2010}, \citet{schaerer2011}, and \citet{cassata2011}, yielding similar trends up to $z\sim 6$. More recent studies confirm these trends (\citealt{jiang2013a,jiang2016,zheng2014}). Simulations likewise predict such correlations (see \citealt{shimizu2011}). However, regardless of the methodology, the scatter in these correlations is non-negligible. Moreover, \citet{atek2014} question the existence of the correlation in their sample. We argue that the scatter observed in the $W_{Ly\alpha}$ dependence on M$_{UV}$/SFR$_{obs}$ can be a consequence of the role played by dust. Galaxies brighter in the UV naturally have a greater Ly$\alpha$ photon production, but the anti-correlation with M$_*$ affects the escape fraction (probably through increased dust extinction). In this scenario, Ly$\alpha$ photon escape is a complex process simultaneously ruled by different properties of high-redshift galaxies. We explore such property space in our analysis of the $W_{Ly\alpha}$ distribution dependence on the M$_{UV}-\beta$ relationship in Section \ref{5.5}.  

Implications involving the fraction of detections as a function of M$_{UV}$ are not straightforward. In principle, this fraction seems to correlate with the UV luminosities of our galaxies, as opposed to what is observed for characteristic $W_{Ly\alpha}$. Nevertheless, our Ly$\alpha$ measurements are flux limited. Therefore, our detection completeness in $W_{Ly\alpha}$ is higher for brighter objects, leading to biases difficult to account for, as noted in \citet{nilsson2009}. Under these circumstances, the ideal approach is to consider both detections and non-detections, while taking into account the uncertainties for line and continuum fluxes ($F$ and $f_\lambda$, respectively). Hence, we encourage further interpretations of these results to focus on the analysis performed on the M$_{UV}-\beta$ plane (Section \ref{5.5}).

\begin{figure}
	\centering
	\includegraphics[width=3.4in]{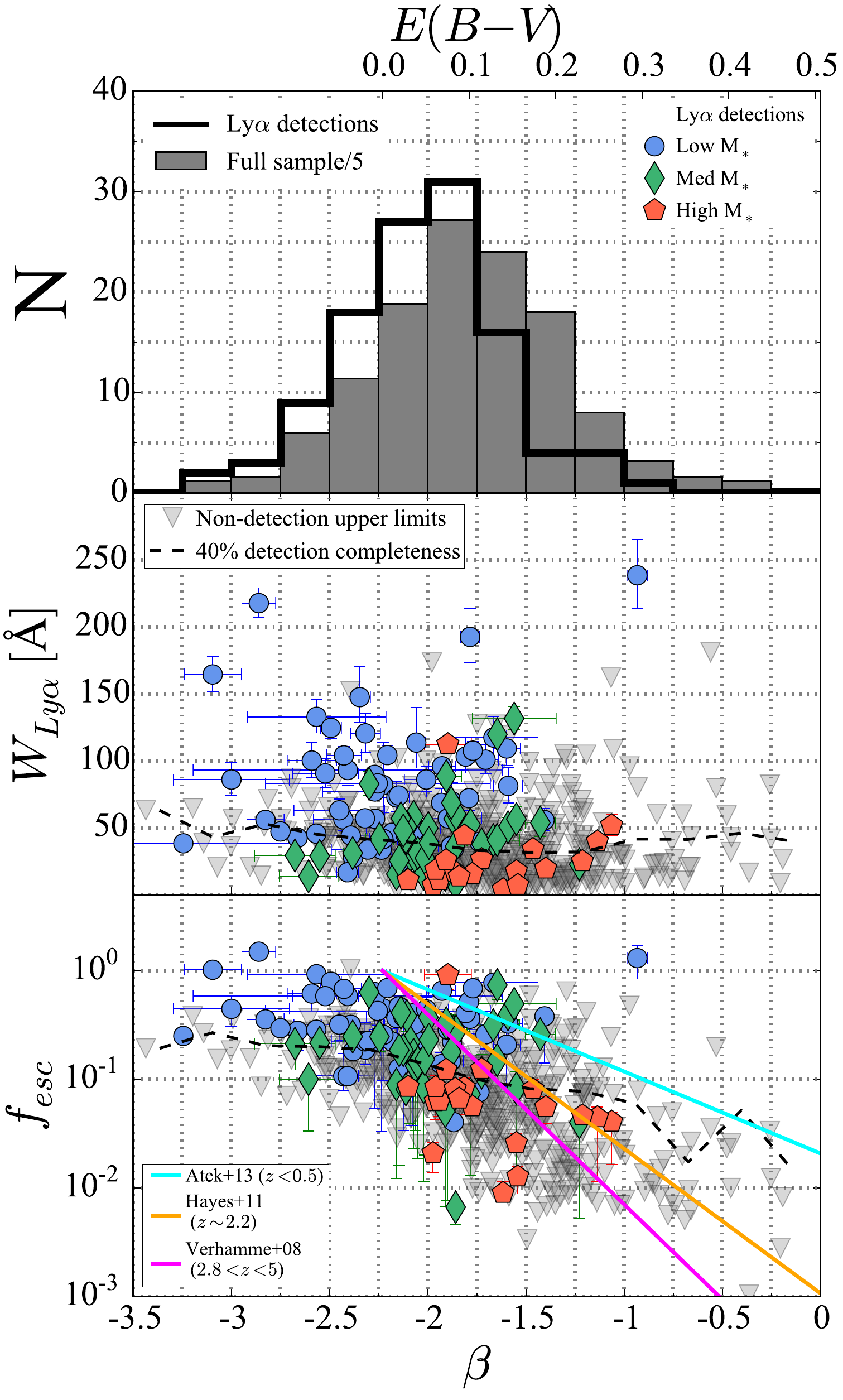}
	\caption{Top: histograms of the sample (gray) and detections (black) as a function of UV slope ($\beta$) and reddening $E(B-V)$. For better comparison, we divide the histogram values for the full sample by five. Center: $W_{Ly\alpha}$ dependence on $\beta$. Bottom: $f_{esc}$ dependence on $\beta$. We are able to measure the UV slope for 611 of the 625 objects composing our sample, and obtain $E(B-V)$ assuming a \citet{reddy2015} attenuation law with a pristine slope of $\beta=-2.23$ (\citealt{meurer1999}). Blue circles, green diamonds, and red pentagons symbolize our low-mass, medium-mass, and high-mass detections, respectively. Gray triangles represent our upper limits from non-detections, which we use to estimate our 40\% completeness (dashed). Our results clearly show that high-$f_{esc}$ objects are associated with extremely blue slopes, consistent with the notion that dust boosts the absorption of Ly$\alpha$ photons. We also include some trends from the literature for comparison.}
	\label{fig:13}
\end{figure}

\subsection{UV slope}
\label{5.4}
Measurement of the UV slopes of high-redshift galaxies is a direct way of tracing the amount of dust inside galaxies, given the assumption of an extinction law and an intrinsic spectral shape. This is of particular interest for Ly$\alpha$ surveys, since simulations (e.g., \citealt{verhamme2008}) and observations at low redshift (\citealt{hayes2011,atek2014}) suggest that dust plays an important role in the escape of Ly$\alpha$ photons. Since we have rest-frame UV photometry from CANDELS for all our objects, we can determine their UV slopes and study their effect on Ly$\alpha$ emission at $z\sim 4$. In this section, we detail our method to estimate the UV slopes for our objects and show our results on the Ly$\alpha$ dependence on this galaxy property. 

To determine UV slopes, we fit a power law $f_\lambda= f_0\lambda^{\beta}$ (\citealt{calzetti1994}) to the photometry of each object. For fitting, we just use a standard least-squares routine on the photometry between rest frame 1400 and 3500\AA. These calculations correspond, in principle, to 8-15 bands between 1500 and 3600\AA. However, since we only use fluxes with S/N$>3$, our median number of bands is eight. Naturally, we further require at least two bands to associate a slope to our targets, which means we can measure the UV slope for 611 of the 625 observed objects composing our sample. 

Our FAST outputs include dust extinction in the $V$-band, A$_V$, for every object. However, if we want to associate a reddening $E(B-V)$ with every galaxy, we need to use our derived UV slopes. We use the relation
\begin{gather}
	\label{ebv}
	f_\lambda= f_0\lambda^{\beta} \propto \lambda^{-2.23}10^{0.4k_{\lambda}E(B-V)}
\end{gather}
where we assume a pristine slope of $\beta=-2.23$ (\citealt{meurer1999}). We also require the adoption of an attenuation law $k_{\lambda}$ (e.g. \citealt{calzetti2000}). \citet{reddy2015} recover a more appropriate attenuation law than \citet{calzetti2000} for high-redshift galaxies, so we use the $k_{\lambda}$ of the former. It is worth noting, nevertheless, that both attenuation laws yield almost identical results.

Since our targets are M$_*$ selected, we have a higher contribution of massive objects in comparison to the M$_*$ distribution of the galaxy population. As more massive objects tend to have higher $E(B-V)$ and redder UV slopes, we expect $z\sim 4$ samples representative of the galaxy population to have a lower contribution from such galaxies. In any case, we show in Figure \ref{fig:13} the UV slope histogram of our sample and our Ly$\alpha$ emitters. Just by comparing both distributions, it is clear that the fraction of emitters increases toward bluer galaxies. We present a quantitative analysis of this claim in Section \ref{5.5}.

We also show in Figure \ref{fig:13} our results on $W_{Ly\alpha}$ and $f_{esc}$ as a function of $\beta$ and $E(B-V)$. Since lower mass galaxies have bluer UV slopes than more massive ones, the trends we find are complementary to our previous results. There is a correlation between the steepness of the UV spectrum and $W_{Ly\alpha}$/$f_{esc}$, although with significant scatter. As extinction seems to play a major role in Ly$\alpha$ photon escape from galaxies, mainly through scattering and absorption (\citealt{blanc2011,hagen2014}), these correlations come as no surprise. Qualitatively, our results agree with the measurements from \citet{shapley2003}, \citet{pentericci2009}, \citet{blanc2011}, and \citet{atek2014}. Regarding the scatter we find at fixed $\beta$ (see also \citealt{blanc2011}), it is consistent with a scenario where the observed Ly$\alpha$ flux is mostly affected by the dependence of the dust-covering fraction on the line of sight. We discuss this picture in detail in Section \ref{5.5}.

When comparing the more dusty Ly$\alpha$ emitters in our sample with results from the literature, however, some differences show up. The $1.9<z<3.8$ survey from \citet{blanc2011} and the $z<0.5$ study from \citet{atek2014} find Ly$\alpha$ emitters up to $E(B-V)\sim 1$. Similarly, \citet{mathee2016} find a population of dusty LAEs at $z\sim 2.23$, and they speculate on how dusty gas outflows might be the feature driving the escape of Ly$\alpha$ radiation. Ly$\alpha$ sources up to $z\sim3$ are \textit{Herschel} (\citealt{oteo2011,oteo2012,casey2012,sandberg2015}) and SCUBA (\citealt{geach2005,hine2016}) detected. Moreover, Ly$\alpha$ emission has also been measured in sub-millimeter galaxies (e.g. \citealt{chapman2005}). On the other hand, our 625 object sample features $\sim5$ objects with $E(B-V)>0.4$, but none of them qualifies as a detection. We cannot state whether the absence of such LAEs in our sample is representative of $z\sim 4$. True enough, the fraction of dusty galaxies at $z\sim 4$ is expected to be lower than at $z<4$. Along such lines, cosmic evolution in the fraction of dusty LAEs has already been discussed in the literature. \citet{blanc2011} study any dust evolution in their $1.9<z<3.8$ LAEs sample and find no significant trend. \citet{hagen2014} study the same sample, and show that there is little anti-correlation, if any, between $E(B-V)$ and redshift. \citet{shapley2003} do observe such evolution in the range $2<z<3.5$, but question the validity given the selection effects associated with LBG surveys. We conclude that even though Ly$\alpha$ emitting galaxies are mostly low-dust objects (e.g. \citealt{song2014}), there is also a population of dustier, low-$W_{Ly\alpha}$ LAEs. Still, the significance of their numbers at $z\gtrsim4$ is still an open question.

Our results also confirm at $z\sim 4$ a trend in $f_{esc}$ that has already been observed at lower redshift. \citet{atek2014} perform a $f_{esc}$ study at $z<0.5$, and recover a similar anti-correlation to the one we show in Figure \ref{fig:13}. \citet{hayes2011} at $z\sim2.2$, \citet{blanc2011} at $1.9<z<3.8$, \citet{song2014} at $z\sim2.1-2.5$, and \citet{mathee2016} at $z\sim 2.23$ also observe the same trends. \citet{verhamme2008} replicate such trends using radiative simulations of galaxies in the range $2.8<z<5$, confirming that qualitative explanations for these observational relations are well supported by theory. We show several best-fit relations from the literature in our lower plot of Figure \ref{fig:13}. Considering that our results are dominated by upper limits, the two higher redshift relations are roughly consistent with our measurements. Still, of particular interest might be the potential redshift evolution suggested by these relations. There is a clear decrease in $f_{esc}$ at high dust contents when going from low to high redshift. If real, this trend could back our previous analysis on the fraction of more dusty LAEs and their evolution as a function of cosmic time. At higher redshift (\citealt{verhamme2008}; this work), low-$W_{Ly\alpha}$, very dusty LAEs do not seem to be common, driving the $E(B-V)-f_{esc}$ relation down significantly for $E(B-V)>0.2$. However, at lower redshift, such objects are actually observed, driving the relation up for $E(B-V)>0.2$.

\begin{figure*}
	\centering
	\includegraphics[width=7in]{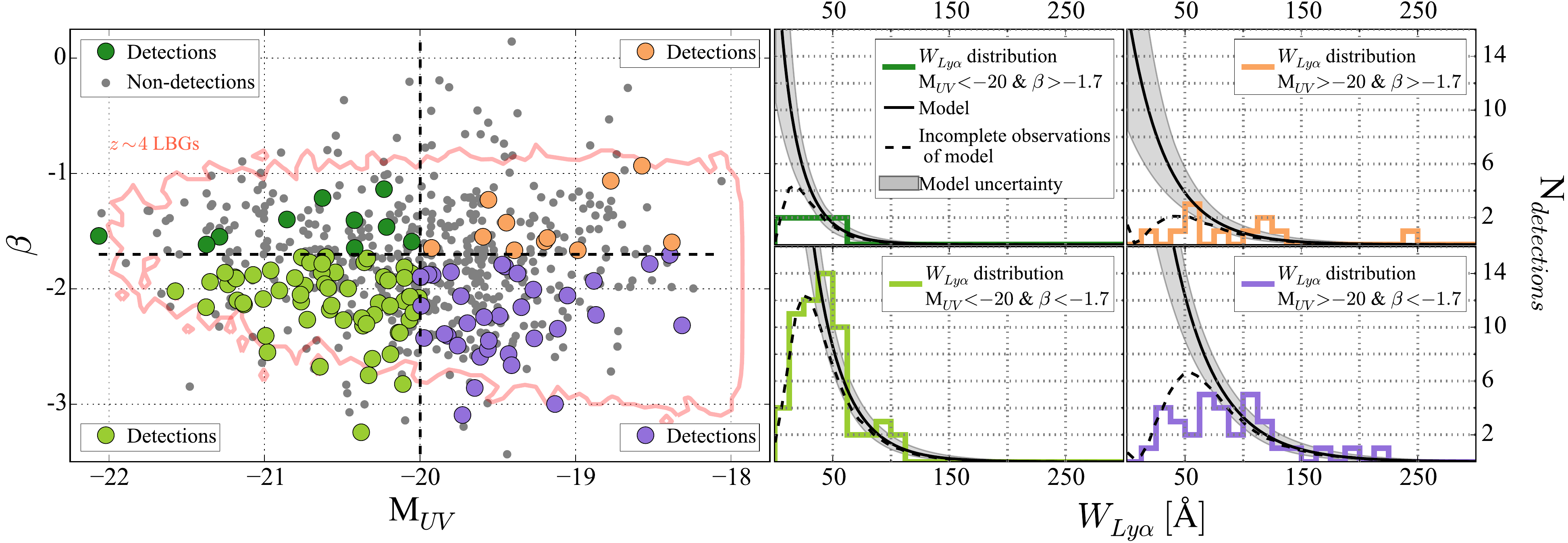}
	\caption{Left: location of our objects in the M$_{UV}-\beta$ plane. Galaxies with detections are plotted as colored circles, whereas non-detections are shown as gray points. Only galaxies for which we can measure M$_{UV}$ and $\beta$ are considered (607 of 625). To show the dependence of $W_{Ly\alpha}$ on this plane, we divide our sample into the four subsets of virtually the same number of galaxies that are delimited by the dashed black lines. We plot in red the 3$\sigma$ contours associated with the relation followed by $z\sim 4$ LBGs, according to the UV LFs from \citet{bouwens2015} and the M$_{UV}-\beta$ relations from \citet{bouwens2014}.  Right: observed $W_{Ly\alpha}$ distributions for the respective M$_{UV}-\beta$ subsets. We plot as black lines the median distributions for each subset given by the best solution of our M$_{UV}-\beta$ model (Equations \ref{Afinal} and \ref{W_0final}). The dotted lines are recovered after applying the incompleteness of every subsample. The shaded contours represent 1$\sigma$ constraints according to our Monte Carlo simulations of Equations (\ref{A}) and (\ref{W_0}).}
	\label{fig:14}
\end{figure*}

\subsection{M$_{UV}$-$\beta$ sequence}
\label{5.5}
We have characterized in this paper the dependence of Ly$\alpha$ emission on M$_*$, SFR, UV luminosity, and UV slope. We find $W_{Ly\alpha}$ and $f_{esc}$ to anti-correlate with these four properties. However, these results might not be independent, since these properties are correlated (see Figure \ref{fig:5}). For instance, M$_*$ and SFR are known to follow a relation at high redshift referred to as the main sequence (\citealt{keres2005,finlator2007,noeske2007,daddi2007}). Similarly, a relation between M$_{UV}$ and $\beta$ has been studied in LBGs (\citealt{bouwens2009,bouwens2014}). As we discuss throughout this work, Ly$\alpha$ escape from galaxies is likely to be a process ruled by many parameters, such as age of the population, gas column density, extinction, and SFR. Therefore, it is interesting to explore Ly$\alpha$ emission dependence on multi-dimensional spaces. In this section, we study $W_{Ly\alpha}$ dependence on M$_{UV}$ and $\beta$. There are two reasons to justify exploring this parameter space in particular. First, both properties are observables, i.e., the amount of assumptions involved in their calculation is kept to a minimum (unlike, for example, M$_*$ and SFR). Second, these two properties are directly related to elements that rule Ly$\alpha$ escape. The UV luminosity of a galaxy traces its ionizing output and gas content, whereas the UV slope is a proxy for dust. Hence, in this section, we focus on characterizing Ly$\alpha$ emission in the M$_{UV}-\beta$ plane. Similarly to previous analyses in this paper, we take advantage of a Bayesian approach to obtain our results.

We first present in Figure \ref{fig:14} the location of our detections and non-detections location in this plane. In qualitative terms, our detections sample most of the sequence initially covered by our targets. For further insight into how our observations depend on this sequence, we construct four subsamples for visualization. The anti-correlation we observe between $W_{Ly\alpha}$ and M$_{UV}$ is still clearly observed when comparing the low M$_{UV}$ with the high M$_{UV}$ subsamples, independent of $\beta$. For the UV slope, however, the trend between $W_{Ly\alpha}$ and $\beta$ does not seem that clear anymore. The normalization of the distribution, however, looks to be highly dependent on the UV slope. To characterize the significance of these insights, we simultaneously model the exponential profile parameters of Equation (\ref{exponential}) as a function of M$_{UV}$ and $\beta$. We start again with linear expressions of the form
\begin{gather}
	\label{A}	
	A(M_{UV}, \beta)= 
	A_{M_{UV}} M_{UV}+ A_{\beta} \beta+A_{C} \\
	\nonumber
	\\
	\label{W_0}	
	W_0(M_{UV}, \beta)= 
	W_{M_{UV}} M_{UV}+ W_{\beta} \beta+W_{C} \\
	\nonumber
\end{gather}
\begin{figure*}
	\centering
	\includegraphics[width=7in]{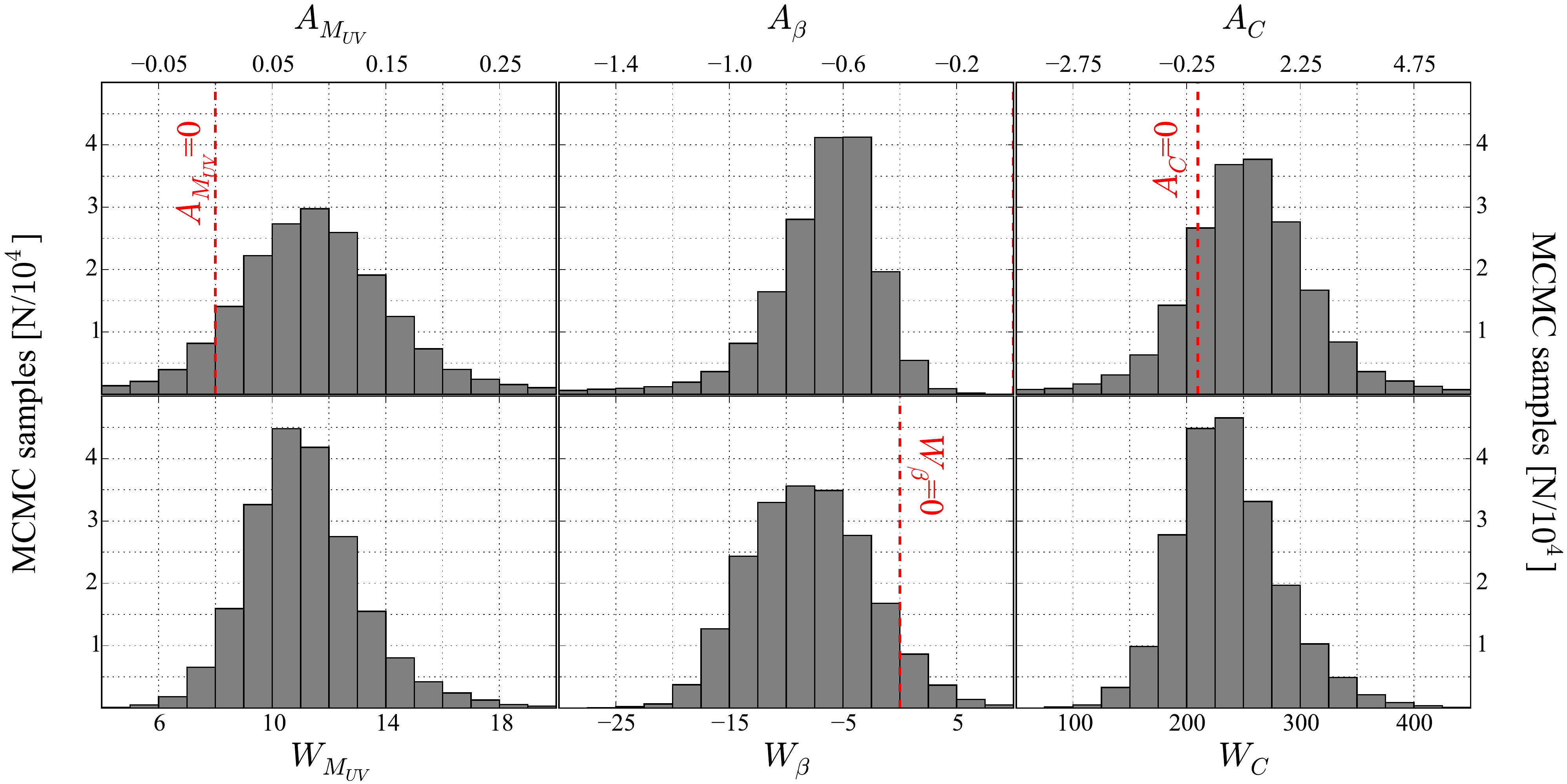}
	\caption{Results from our Monte Carlo simulation on the dependence of Ly$\alpha$ emission on the M$_{UV}-\beta$ plane. The six parameters correspond to the coefficients in Equations (\ref{A}) and (\ref{W_0}). The histograms are obtained after collapsing the posterior distribution for every parameter. Left: $A, W_0$ dependence on M$_{UV}$. Center: $A, W_0$ dependence on $\beta$. Right: additive constants for $A$ and $W_0$. These results suggest that the normalization of the distribution is mostly ruled by dust, whereas the $W_{Ly\alpha}$ is mostly determined by UV luminosity (see Figure \ref{fig:14}).}	
	\label{6posterior}
	\label{fig:15}
\end{figure*}
Once the parameterization and priors are set, we can obtain the posterior distribution of $\{W_n\}=(A_{M_{UV}}, A_{\beta}, A_C, W_{M_{UV}}, W_{\beta}, W_C)$ using Equation (\ref{final}). We assume the priors to be independent, which translates to $p(\{W_n\})\propto \displaystyle \prod p(W_i)$. Once again, we impose ignorance on the parameters, i.e., we adopt non-informative priors. We study the posterior distribution assuming parameter ignorance in both linear and logarithmic scales. We decide for the first, since logarithmic priors diverge for parameters than can adopt values close to zero ($A_{M_{UV}}$, $A_{C}$ and $W_{\beta}$). Therefore, our prior is simply a constant $p(\{W_n\})$. Considering that $p(\{F\})$ is just a normalization factor in Equation (\ref{final}), the posterior distribution of the six-parameter space can be obtained:
\begin{gather}
	p(W_n| \{F\})\propto \displaystyle \prod_{D}p(F_{i}|W_n) \displaystyle \prod_{ND}p(F_{i}<F^{*}_{i}|W_n)
\end{gather}

The value of the parameters $A$ and $W_0$ is, by definition, constrained to the intervals $A=[0,1]$ and $W_0=(0,\infty)$. Still, in the case of extreme luminosities and/or slopes, our linear parameterizations can yield values outside these intervals. As a solution, we just impose $A$ and $W_0$ to saturate outside their corresponding ranges. This condition can be considered simply as an indirect prior on the parameters. For the particular case of $W_0<0$, we impose $A=0$, i.e., the $W_{Ly\alpha}$ distribution is a Dirac delta (i.e., no Ly$\alpha$ emission and/or absorption). 

We recover our best solution from Monte Carlo simulations and obtain the uncertainties on our parameters using the collapsed distributions. The results are as follows:
\begin{gather}
	\nonumber
	A(M_{UV}, \beta)= \\
	\label{Afinal}	
	0.08^{+0.1}_{-0.06} M_{UV}-0.6^{+0.1}_{-1.2}\beta+1.1^{+1.5}_{-1.4} \\
	\nonumber
	\\
	\nonumber
	W_0(M_{UV}, \beta)= \\
	\label{W_0final}	
	11^{+2.0}_{-1.8}M_{UV}-7.8^{+5.2}_{-5.5}\beta+235^{+45}_{-42}\\
	\nonumber
\end{gather}
We show with more detail the results for these coefficients in Figure \ref{fig:15}. The early analysis based on the sample binning from Figure \ref{fig:14} is verified in Figure \ref{fig:15}. The normalization factor $A$ anti-correlates with $\beta$ with a significance $\gtrsim2\sigma$. There also seems to be an anti-correlation between $A$ and M$_{UV}$, but to a level of $\sim 1\sigma$. For $W_{0}$, the e-folding scale of the distribution, the behavior is the opposite. The extent of the distribution correlates with M$_{UV}$ and shows a weak anti-correlation with $\beta$.

We now focus on the interpretation of these results. The fact that beta correlates mostly with $A$ suggests that the UV slope is somehow related to a stochastic probability of being able to observe or not a particular object in Ly$\alpha$. On the other hand, the stronger correlation of M$_{UV}$ with $W_{Ly\alpha}$ suggests that the UV luminosity is associated with physical processes that determine the magnitude of the resulting $W_{Ly\alpha}$. A possible scenario is one in which M$_{UV}$ traces SFR and, therefore, the total cold gas mass of galaxies. Hence, higher UV luminosities translate into increased levels of Ly$\alpha$ photon scattering into an extended halo, smoothly reducing the $W_{Ly\alpha}$ of the central source. At the same time, in a significantly clumpy ISM in which dust is well mixed with the gas, a large contrast in gas density/column will imply that dust can effectively block the bulk of both UV and Ly$\alpha$ radiation in some parts of the disk. Such impact will be significant toward certain lines of sight, with little to no effect on others. Such ``covering factor" scenario could explain that $\beta$ correlates more significantly with the normalization ($A$) than with the shape of the distribution ($W_{0}$). This is consistent with the finding that Ly$\alpha$ and UV suffer from similar levels of extinction by dust in LAEs (e.g. \citealt{blanc2011}). Moreover, the scatter in the $f_{esc}$ of Ly$\alpha$ photons at fixed reddening (see Figure \ref{fig:13} and \citealt{blanc2011}) is also consistent with this scenario where wide ranges of dust absorption lines of sight and photon scattering halos rule the escape of Ly$\alpha$ radiation.


Naturally, this interpretation is a simplified description of Ly$\alpha$ escape from galaxies. Going no further, Figure \ref{fig:15} reveals how both distribution parameters do not solely depend on one property. Ideally, studies of $W_{Ly\alpha}$ and $f_{esc}$ in three-parameter spaces (e.g. M$_*$, M$_{UV}$, $\beta$) can yield predictive and more accurate parameterization of Ly$\alpha$ emission. However, such analysis must be performed in much larger datasets than ours, at least if significant enough results are to be obtained. Furthermore, hydrodynamical simulations can also give insight into how Ly$\alpha$ escape depends on the line of sight and the distribution of gas, dust, and star-forming regions (e.g. \citealt{verhamme2012}), especially if performed in a statistically significant sample.

\begin{figure*}
	\centering
	\includegraphics[width=7in]{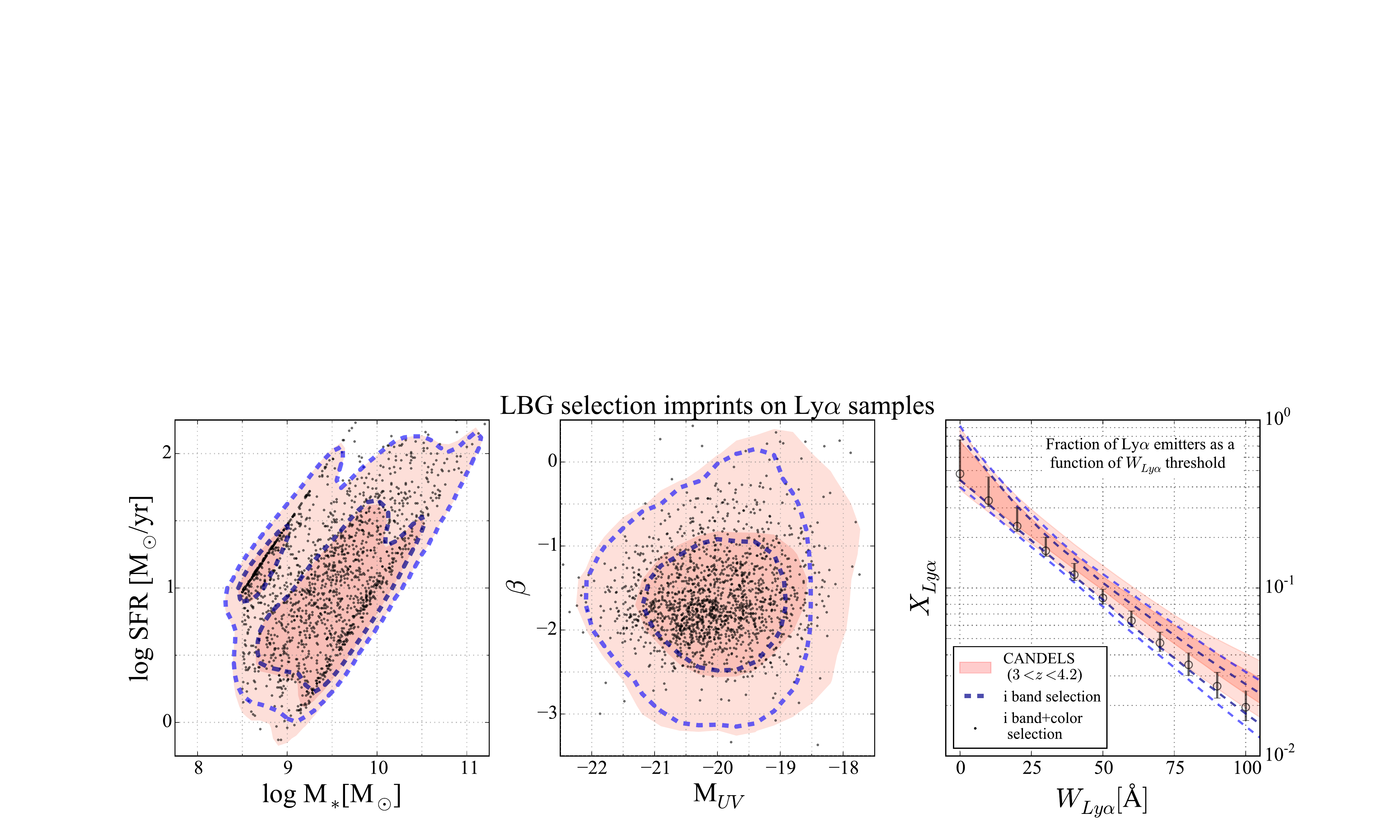}
	\caption{Selection biases induced by LBG surveys. We show in red all $3<z<4.2$ CANDELS galaxies. Blue contours represent galaxies that satisfy $z\sim 4$ LBG detection band selections ($V_{606}>5\sigma$). LBGs (detection band and color selections) are shown in black, with opacity a proxy for number density. To avoid incompleteness in these samples, we only consider CANDELS objects with M$_*>10^{8.5}\mbox{M}_{\odot}$. Left: M$_*$-SFR sequence for the three samples (contours represent 1$\sigma$ and 2$\sigma$ confidence regions). Center: M$_{UV}-\beta$ sequence for the three samples (1$\sigma$, 2$\sigma$). Right: fraction of Ly$\alpha$ emitters with equivalent width higher than $W_{Ly\alpha}$ for each sample (1$\sigma$ and 2$\sigma$ contours; 1$\sigma$ error bars). We do not observe any major biases in LBG samples besides unavoidable detection band limitations. If anything, the color selection $(B-V>1.1)$ removes low-$\beta$ galaxies, lowering the Ly$\alpha$ fraction at all $W_{Ly\alpha}$. We remark that this analysis is only valid at $z\sim4$ for galaxies with M$_*>10^{8.5}\mbox{M}_{\odot}$.}	
	\label{fig:16}
\end{figure*}

\section{Ly$\alpha$ Dependence on Sample Selection}
\label{6}
\subsection{LBG Samples}		
\label{6.1}	
The Lyman break selection technique has proven to be a very efficient method for detecting high-redshift galaxies (e.g. \citealt{steidel1996,shapley2003,stark2010,ono2012}). The fact that the Lyman break is in the optical region of the observed spectrum for galaxies at redshift $z\geqslant3$ allows for efficient detection from ground telescopes. By only requiring the use of a few broadband filters, several galaxies that can be detected in a deep single exposure. Still, to avoid aliasing with the Balmer break, there are unavoidable biases associated with this technique. First, galaxies with no prominent Lyman break are, by construction, not detected. As a consequence, either extremely young or passive, heavily extincted galaxies are underrepresented in LBG surveys. Second, these surveys also impose color restrictions on the slope of galaxies, further increasing the selection toward, in principle, bluer UV objects. Third, this technique is limited by the M$_{UV}$ sensitivity of the survey, creating incompleteness at low-SFR. In the case of Ly$\alpha$ emission, this observational limit can have a significant downside. As shown in Section \ref{5.2}, there is a clear anti-correlation between UV luminosity and $W_{Ly\alpha}$, leading to the possibility that Ly$\alpha$ studies in LBG samples are missing the highest $W_{Ly\alpha}$ galaxies in the universe. Indeed, the deepest surveys nowadays reach $\sim 50\%$ completeness at observed m$_{UV}\sim29.5$ (e.g \citealt{bouwens2015,bowler2015,bowler2017,ishigaki2017}). These limits translate to higher redshift $\sim 50\%$ completeness at M$_{UV}\sim$ -18.3, -19, -19.5, and -19.7 ($z\sim$ 4, 5, 6, and 7, respectively). These limitations become problematic when comparing the results between LBG and narrowband samples. Presumably due to not requiring a continuum detection, the $W_{Ly\alpha}$ observed in narrowband surveys are systematically higher (e.g. \citealt{zheng2014}).

Our dataset and results can be useful for characterizing the effects that selection techniques can have on Ly$\alpha$ emission. Even though it is possible to identify these selections directly in our sample, any comprehensive analysis must consider our galaxy selection procedure. Since our sample follows neither the M$_*$ or M$_{UV}$ functions of the galaxy population, correcting requires assumption of M$_*$ and/or M$_{UV}$ distributions. Nevertheless, we can still simulate Ly$\alpha$ emission samples on CANDELS galaxies using our results of $W_{Ly\alpha}$ dependence on the M$_{UV}-\beta$ plane given by Equations (\ref{Afinal}) and (\ref{W_0final}). In this section, we describe our simulation of a $z \sim 4$ LBG sample from the 3D-HST catalogs. We then compare the properties of LBGs with the parent distribution, focusing on M$_*$, SFR, M$_{UV}$, and $\beta$. We finally simulate Ly$\alpha$ fractions for each sample and conclude on the effects of LBG selection at $z \sim 4$.

LBGs at $z\sim 4$ are typically selected using B-dropouts and imposing color selections on redder filters. As we are simulating this survey in CANDELS, our detection limits are given by the depth of their $H_{160}$ images. For the $z\sim 4$ LBG selection, we adopt the same methodology applied in \citet{bouwens2012}:
\begin{gather}
	\nonumber
	\mbox{B-dropouts:	}	(B_{435}-V_{606}>1.1)\  \wedge (B_{435}-V_{606} \\
	\label{color}
	>(V_{606}-z_{850})+1.1) \wedge (V_{606}-z_{850}<1.6).
\end{gather}
After color selection, we impose $5\sigma$ detections in the $V_{606}$ filter, to which we now refer to as the detection band. As commonly applied in LBG surveys, we also require all candidates to have $<1.5\sigma$ measurements in the $U$ band, since the Lyman break must be redshifted out of this filter. Before performing any analysis on the CANDELS LBG sample, we run EAZY and FAST according to our prescriptions (see Section \ref{2.3}). Of the $\sim 3200$ CANDELS galaxies that comply with our selection, $\sim 800$ classify as low-redshift interlopers ($z_{EAZY}<3$), according to our outputs. We check EAZY 3$\sigma$ constraints for these interlopers and find most to have $z_{99}<4$ . Hence, we find a low-redshift contamination in $z\sim 4$ CANDELS LBGs that is higher than the typically reported number of $\sim 10 \%$ (\citealt{bouwens2007}). From now on, we remove these presumed $z_{EAZY}<3$ contaminants and work with a $z\sim 4$ LBG sample of nearly 2300 objects.

Using our outputs, we plot relevant properties of CANDELS LBGs in Figure \ref{fig:16}. To ensure that we do not venture way beyond CANDELS completeness, we restrict the upcoming analysis to galaxies with M$_*>10^{8.5}\mbox{M}_{\odot}$. We construct two $3<z<4.2$ CANDELS samples for comparison: all galaxies with M$_*>10^{8.5}\mbox{M}_{\odot}$ and all $V$-band detected M$_*>10^{8.5}\mbox{M}_{\odot}$ objects. As seen in the central panel, the major LBG selection effect is associated with the detection band threshold. These left-out objects are either red, heavily extincted galaxies, or blue, intrinsically faint objects (\citealt{quadri2007}). Indeed, as revealed by the comparison between LBGs and the detection-band-selected sample, any selection imprints associated to color requirements are minor. It is still worth noting, however, that these color criteria seem to neglect blue rather than red galaxies at $z\sim 4$ (center of Figure \ref{fig:16}). We find the $(B-V>1.1)$ Lyman break cut to be the driving criterion behind this selection effect (see Equation \ref{color}). We remark that these insights might not be valid at different redshifts or lower M$_*$, however. Some studies have indeed explored differences at higher redshift (e.g. \citealt{jiang2013a,jiang2013b,jiang2016}). In analyses that explore Ly$\alpha$ emission strength, UV continuum properties, morphologies, and sizes, they find no significant differences between LBGs and LAEs at $z\geqslant 6$. 

The insights we present can be further tested by simulating the fraction of LAEs above an $W_{Ly\alpha}$ threshold. We use our results from Section \ref{5.5} to estimate the Ly$\alpha$ fraction for each sample and plot our results in the right panel of Figure \ref{fig:16}. Indeed, depending on the depth of the detection band, the fraction of low UV luminosity galaxies composing the sample can vary significantly. As a consequence, deeper surveys can potentially recover a higher fraction of line emitters at high $W_{Ly\alpha}$. To deal with this selection, it has been proposed to compare the Ly$\alpha$ fractions over narrow UV luminosity samples (\citealt{stark2011,mallery2012,ono2012,schenker2012}). Our results in this paper further emphasize the need to compare galaxies of the same luminosity and properly characterize completeness when studying the evolution of the Ly$\alpha$ fraction with cosmic time. Regarding color selections, we find the Ly$\alpha$ fraction to be slightly lower for LBGs than for the $V$-band-detected sample. As we suggested, this is a consequence of LBG color cuts leaving out some of the bluest galaxies.

\begin{figure}
	\centering
	\includegraphics[width=3.4in]{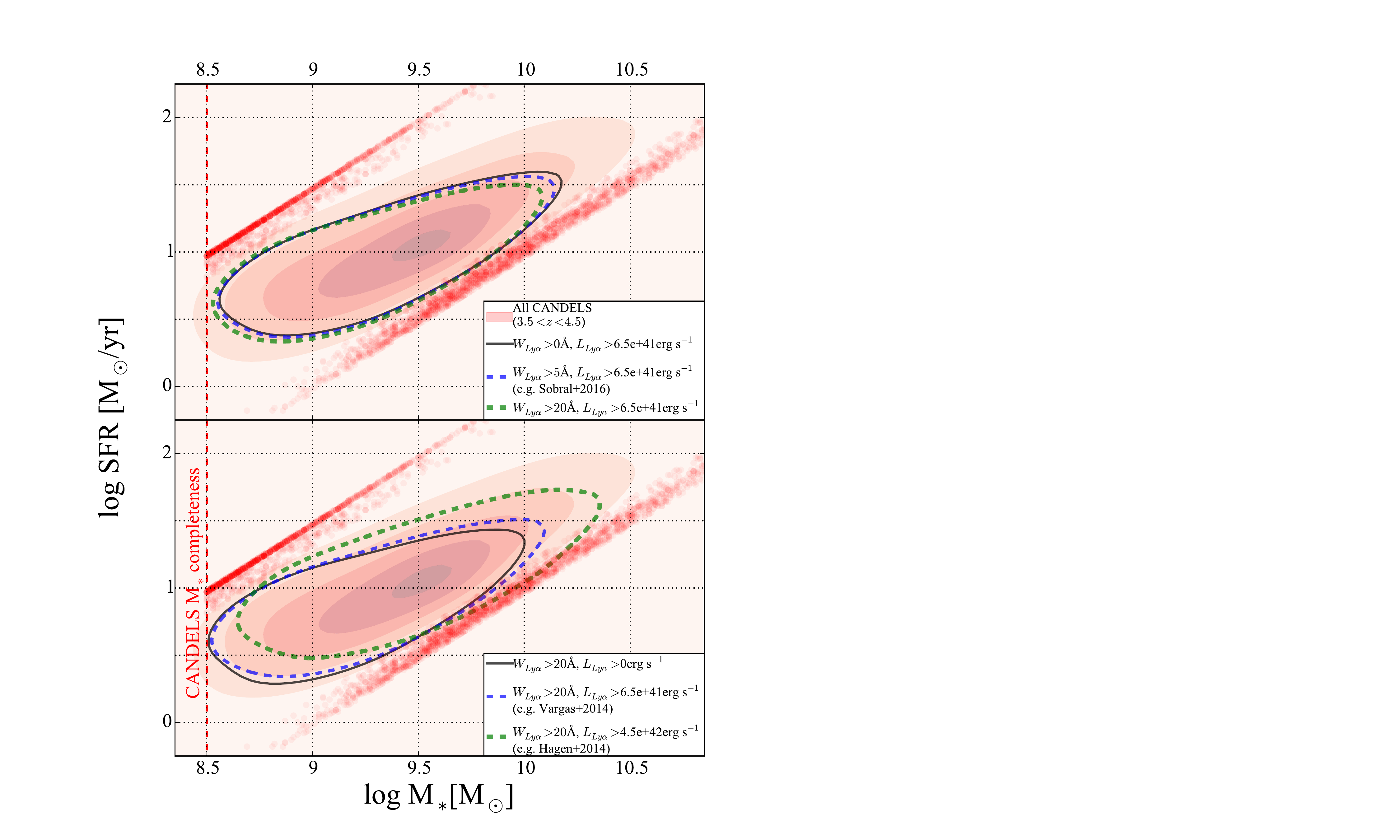}
	\caption{Selection effects produced by narrowband-selected samples of LAEs. For the parent sample, we show as red contours the M$_*$-SFR sequence of $3.5<z<4.5$ CANDELS galaxies with M$_*>10^{8.5}\mbox{M}_{\odot}$. We adopt this value as the threshold for which CANDELS at $z\sim 4$ is fairly complete in GOODS-S (\citealt{duncan2014}). For clarity, we also remove CANDELS galaxies in the young and old age tracks (red datapoints). Top: effect of $W_{Ly\alpha}$ selections on LAE samples. We simulate the resulting LAE samples under no $W_{Ly\alpha}$ selections (black), $W_{Ly\alpha}>5$\AA \ (\citealt{sobral2016}; blue), and $W_{Ly\alpha}>$ 20 \AA \ (green). $W_{Ly\alpha}$ cuts are more likely to remove high M$_*$, high-SFR LAEs, but the effect for typical survey thresholds ($W_{Ly\alpha}\leqslant 20$) is minor. Bottom: effect of Ly$\alpha$ luminosity ($L_{Ly\alpha}$) selections on LAE samples. We simulate the region sampled by the narrowband-selected surveys described in \citet{vargas2014} and \citet[blue and green, respectively]{hagen2014}. We predict the deeper observations of \citet{vargas2014} to statistically measure lower-SFR LAEs than \citet{hagen2014}. Hence, luminosity selections remove mostly low-SFR objects. The flux limitations of these surveys have been translated to luminosity in order to be simulated at $z\sim 4$.}
	\label{fig:17}
\end{figure}

\subsection{Narrowband Samples}
\label{6.2}
Ly$\alpha$-emitting galaxies can also be selected using narrowband imaging or blind spectroscopy (e.g. \citealt{gronwall2007, ouchi2008, adams2011}). By establishing a $W_{Ly\alpha}$ detection threshold between the narrow- and broadband flux measurement, this technique allows for efficient line-emitter selection. Since only line detection is required, such surveys can trace fainter objects than the LBG technique, possibly leading to selection of the youngest and faintest galaxies at high redshift. However, even though most narrowband measurements of LAEs are followed up by spectroscopy, any sample selection effects induced by the narrowband technique are already present in the sample. The $W_{Ly\alpha}$ threshold used for detection, which is determined by the ability to separate low-redshift interlopers from high-redshift LAEs, can adopt a wide range of values. Depending on redshift, observer frame thresholds translate into different rest-frame $W_{Ly\alpha}$ cuts. For instance, \citet{vargas2014} select sources with $W_{Ly\alpha}>20$\AA \ at $z\sim 2.1$, \citet{zheng2014} select $W_{Ly\alpha}>9$\AA \ at $z\sim 4.5$, and \citet{sobral2016} go as low as $W_{Ly\alpha}>5$\AA \ at $z\sim 2.23$. If $W_{Ly\alpha}$ selections induce important biases on galaxy samples, the comparison of different surveys is not straightforward.

 In this section, we explore the effects that narrowband selections have on the population of LAEs, focusing on the M$_*$-SFR plane. The insights we present here are based on Equations (\ref{Afinal}) and (\ref{W_0final}), i.e., our M$_{UV}-\beta$ model. We show in Figure \ref{fig:17} the outcome of $W_{Ly\alpha}$ (top) and line-flux selections (bottom) on the M$_*$-SFR sequence. The red contours show $3.5<z<4.5$ CANDELS galaxies with M$>10^{8.5}\mbox{M}_{\odot}$. For clarity, however, we concentrate our analysis on the main sequence, removing objects with young and old ages (red dots). This approach allows our results to be dominated by objects optimally fitted by our FAST executions. 
 
To explore the effect of $W_{Ly\alpha}$ selections (top panel of Figure \ref{fig:17}), we base our analysis on the $z\sim 2.23$ narrowband survey of \citet{sobral2016}, whose galaxies are limited to $W_{Ly\alpha}>5$\AA \ and fluxes $>2\times 10^{-17}$ erg cm$^{-2}$ s$^{-1}$ (5$\sigma$). We simulate a flux-limited-only Ly$\alpha$ sample and use it as baseline (black). We then show the region where $W_{Ly\alpha}>5$\AA \ (1$\sigma$; blue) and $W_{Ly\alpha}>20$\AA \ (1$\sigma$; green) objects lie. The $W_{Ly\alpha}>5$\AA \ contours are intended to reproduce \citet{sobral2016} selections, whereas the more restrictive selection $W_{Ly\alpha}>20$\AA \ is representative of multiple surveys (\citealt{gronwall2007,hagen2014,vargas2014}). As confirmed by this plot, galaxies with higher characteristic $W_{Ly\alpha}$ (i.e., low M$_*$, low-SFR) are more likely to be selected by narrowband samples, even though the effect is minor for the typical narrowband cuts of $W_{Ly\alpha}\leqslant20$\AA. Still, selections based solely on $W_{Ly\alpha}$ systematically fail to remove AGNs and neglect bright Ly$\alpha$ emitters (\citealt{sobral2016}). All these insights remark on the importance of using low $W_{Ly\alpha}$ cuts and thorough interloper controls.
 	
After narrowband outcomes are used for sample selection, spectroscopic observations follow. However, these follow-ups of Ly$\alpha$ emitting galaxies are flux limited, just like the ones presented in this work. Still, as we consider completeness in our modeling and simulations of Section \ref{5.5}, we can still make inferences on flux limited studies. We use our Monte Carlo simulation outputs to also assess the effects of line luminosity selections. Using the corresponding M$_*$, SFR, and $f_{UV}$ for every object, we obtain the probability of $L_{Ly\alpha}>L_{Ly\alpha}^*$ for every galaxy. The outcome (bottom panel of Figure \ref{fig:17}) reveals that flux selections bias samples toward high-SFR LAEs.
 
Discrepancies in the location of LAEs in the M$_*$-SFR plane have already been observed in the literature. In Figure 10 of their paper, \citet{hagen2014} plot the M$_*$-SFR relation of their $z\sim 2-3$ LAEs alongside the $z\sim 2.1$ counterparts of \citet{vargas2014}. The \citet{hagen2014} objects are part of the HETDEX Pilot Survey (\citealt{adams2011}), for which detections are constrained to line fluxes $>7-10\times 10^{-17}$ erg cm$^{-2}$ s$^{-1}$ (5$\sigma$) and $W_{Ly\alpha}>20$\AA. In contrast, the \citet{vargas2014} flux depth is higher than that of HETDEX, reaching $2\times 10^{-17}$ erg cm$^{-2}$ s$^{-1}$ (5$\sigma$), while also selecting sources with $W_{Ly\alpha}>20$ \AA. Therefore, \citet{hagen2014} are comparing two $W_{Ly\alpha}$-selected surveys with different line-flux depths. Our results in this section would, then, predict \citet{vargas2014} LAEs to sample lower M$_*$ and SFRs because they go deeper (see $\sim 10^9$ M$_{\odot}$ galaxies in Figure \ref{fig:17}). This is in fact the pattern observed in Figure 10 of \citet{hagen2014}, i.e., we can qualitatively reproduce their comparison with our simulations from the results of Section \ref{5.5}. This explanation is backed by the fact that these discrepancies are not caused by inconsistent M$_*$ or SFR derivations. Both studies assume a \citet{salpeter1955} IMF, cSFHs, and a fixed metallicity of $Z=0.2Z_{\odot}$.

\citet{vargas2014} and \citet{hagen2014} find M$_*<10^9$M$_{\odot}$ LAEs to lie above the M$_*$-SFR relation. A similar trend is reported in the work of \citet{karman2017}, who study Ly$\alpha$ emitters down to $10^6$M$_{\odot}$. \citet{karman2017} point out that this offset can be a consequence of how uncertain SFRs are for low M$_*$, starburst galaxies. This explanation is consistent with the fact that most of our M$_*<10^9$M$_{\odot}$ Ly$\alpha$ emitters are constrained to the lowest age locus of the M$_*$-SFR plane (see Figure \ref{fig:5}). Moreover, \citet{finkelstein2015} observe the same lowest age, low M$_*$ Ly$\alpha$ emitters at $z\sim 4.5$. To answer whether low M$_*$ LAEs lie above starforming galaxies or the slope of the sequence changes toward M$_*<10^9$M$_{\odot}$, different approaches are required. For instance, no discrepancies are found between $z\sim 2$ LAEs and H$\alpha$ sources (see \citealt{mathee2016}), although most of their galaxies have M$_*>10^9$ M$_{\odot}$. In summary, evidence suggests that $z>2$, M$_*<10^9$ M$_{\odot}$ LAEs lie above extrapolations of the M$_*$-SFR relation. It is unclear whether this offset is real or the position of the relation at low M$_*$ is far from certain.

\begin{figure*}
	\centering
	\includegraphics[width=7in]{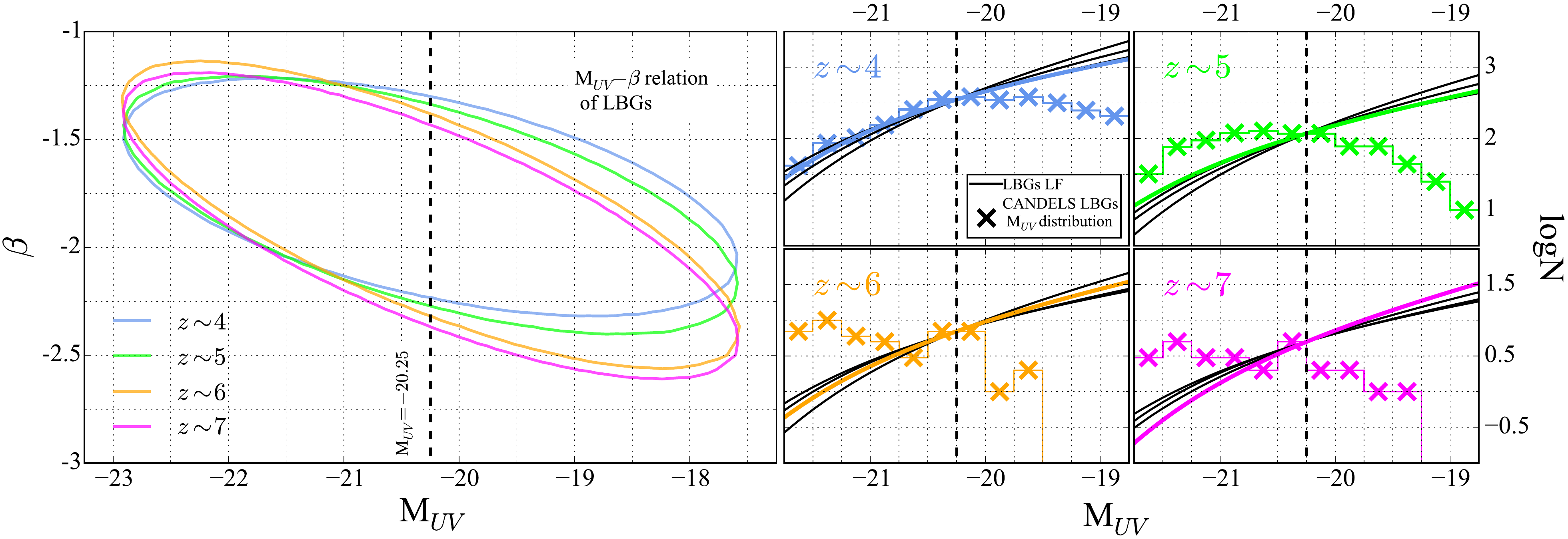}
	\caption{Simulated LBGs that we use to predict the Ly$\alpha$ fraction in LBG samples as a function of redshift. Left: 1$\sigma$ contours of the M$_{UV}-\beta$ relations we use for these $4\leqslant z \leqslant 7$ simulations. We start from the best-fit relations found by \citet{bouwens2014} and then add an intrinsic scatter of 0.35 to the relation (\citealt{bouwens2012, castellano2012, bouwens2014}). Right: M$_{UV}$ distribution of our simulated LBGs. The UV complete samples (solid lines) are drawn following the LFs from \citet{bouwens2015}. The incomplete samples (crosses) are obtained after simulating an LBG selection from CANDELS. The LFs are normalized to the number of CANDELS LBGs at M$_{UV}\sim -20.25$. In order to construct our samples, we first draw from the corresponding M$_{UV}$ distribution (right panel) and then associate a UV slope following the M$_{UV}-\beta$ relation (left panel). For the resulting Ly$\alpha$ fractions, refer to Figure \ref{fig:19}.}
	\label{fig:18}		
\end{figure*}

\section{Inferences on the $4\leqslant \lowercase{z}\leqslant 7$ L\lowercase{y$\alpha$} Fraction}
\label{7}
Probably the most important use for Ly$\alpha$ emission is tracing the neutral hydrogen fraction in the IGM. Several studies over the last decade have constrained the fraction of Ly$\alpha$-emitting galaxies as a function of redshift (\citealt{stark2010,stark2011,ono2012,schenker2012,tilvi2014,cassata2015}), bringing us closer to the goal of constraining the epoch of reionization.  The Ly$\alpha$ fraction is fairly well understood at $z<5$ (\citealt{stark2011,cassata2015}), with most efforts nowadays focusing on $z\sim7,8$ (\citealt{ono2012,treu2013,tilvi2014,furusawa2016}). Along these lines, our characterization of $W_{Ly\alpha}$ on the M$_{UV}-\beta$ plane can be used to simulate $W_{Ly\alpha}$ distributions at higher redshift. In this section, we apply the observed M$_{UV}-\beta$ relations from \citet{bouwens2014} and LFs from \citet{bouwens2015} to simulate high-redshift $W_{Ly\alpha}$ distributions. By means of these simulations, we can predict the Ly$\alpha$ fraction in galaxies up to $z\sim 7$, providing the first semi-analytical constraint to this tracer toward the reionization epoch. Furthermore, we also simulate dropouts from CANDELS photometry to explore the effects of observational limitations on the inferred Ly$\alpha$ fractions at high redshift. 

Our results from this section are based on assuming the same $z\sim 4$ $W_{Ly\alpha}$ dependence on M$_{UV}-\beta$ at every redshift. We remark that testing this assumption with currently available datasets is challenging, since different sample selections and incompleteness levels can affect the observed $W_{Ly\alpha}$ dependence on the M$_{UV}-\beta$ plane. It is also worth noting that the analysis described in this section does not account for any effects related to changes in the merger fraction, IGM opacity (e.g. \citealt{gunn-peterson1965,becker2001,fan2006}), and/or conditions of the ISM with cosmic time (e.g. \citealt{carilli2013}). As the number of complete, unbiased Ly$\alpha$ surveys grows (e.g. \citealt{cassata2015,hathi2016}; this work), we will be able to test these assumptions and explore their dependence on redshift.

We now detail our simulation of high-redshift LBG samples. For the LFs, we use \citeauthor{bouwens2015}'s \citeyear{bouwens2015} best Schechter parameters (excluding CANDELS-EGS). We start by drawing objects following the LF and then associate a UV slope according to the best-fit relations from Table 3 of \citet{bouwens2014}. We finish by adding an intrinsic scatter of 0.35 to the slopes at every redshift (\citealt{bouwens2012, castellano2012, bouwens2014}). Similarly, we perform an analogue procedure following the M$_{UV}$ distribution of CANDELS LBGs, which allows us to assess the effect of magnitude incompleteness. In order to do so, we make use of the optical and IR photometry publicly available from the 3D-HST catalogs (\citealt{skelton2014}). For every redshift, we use the selections from \citet{bouwens2015}, since they are based on CANDELS photometry:
\begin{gather}
\nonumber
\mbox{$z\sim$ 4:	}	(B_{435}-V_{606}>1) \wedge (i_{775}-J_{125}<1)\\
\wedge (B_{435}-V_{606}>1.6(i_{775}-J_{125})+1)
\end{gather}
\begin{gather}
	\nonumber
	\mbox{$z\sim$ 5:	}	(V_{606}-I_{814}>1.3) \wedge 	(I_{814}-H_{160}<1.25)\\
\wedge (V_{606}-I_{814}>0.72(I_{814}-H_{160})+1.3)
\end{gather}
\begin{gather}
	\nonumber
	\mbox{$z\sim$ 6:		} (I_{814}-J_{125}>0.8) \wedge 	(J_{125}-H_{160}<0.4) \\
\wedge (I_{814}-J_{125}>2(J_{125}-H_{160})+0.8)
\end{gather}
\begin{gather}
	\nonumber
	\mbox{$z\sim$ 7:		}(I_{814}-J_{125}>2.2) \wedge 	(J_{125}-H_{160}<0.4) \\
\wedge (I_{814}-J_{125}>2(J_{125}-H_{160})+2.2).
\end{gather}

Similarly to our Section \ref{6.1} selection, we impose S/N cuts in the detection bands. For $z\sim 4$ and $5$ dropouts, we require $3\sigma$ detections in the $V_{606}$, $z_{850}$ bands. For $z\sim 6$ and 7 dropouts, we impose at least $3\sigma$ detections in the $Y_{102}$ and $J_{125}$ bands, respectively. To associate a dropout redshift with every galaxy, however, we use the photometric redshifts from 3D-HST instead of the actual dropout from where it was selected. By doing so, we are in agreement with the methodology of \citet{bouwens2015}. We finally complete these CANDELS LBG samples by associating UV slopes to every galaxy. Just as for the complete sample, we associate the slopes following the M$_{UV}-\beta$ relations from \citet{bouwens2014}.

We present in Figure \ref{fig:18} the assumed M$_{UV}-\beta$ sequences and the UV luminosity distribution for both sets under consideration. Since we are starting from the \citet{bouwens2014}  M$_{UV}-\beta$ characterization, the steepness of the linear relation increases with redshift. The contours we show in Figure \ref{fig:18} (left panel) are obtained after incorporating the intrinsic scatter, which we assume to be of 0.35. Note that the addition of such a scatter does not render the redshift evolution of the M$_{UV}-\beta$ distribution negligible. According to our Equations (\ref{Afinal}) and (\ref{W_0final}), this implies that the probability of observing $W_{Ly\alpha}$ in emission at fixed M$_{UV}$ should increase with redshift. The right panel of Figure  \ref{fig:18} shows the absolute magnitude distribution of $4\leqslant z \leqslant 7$ LBGs as found by \citet[solid lines]{bouwens2015}. Note the growth of the LF as a function of redshift. Since the LF becomes steeper as the redshift increases, the probability of observing a high-$W_{Ly\alpha}$ galaxy should also increase. One more important point regarding Figure \ref{fig:18} must be noted. As initially suggested, CANDELS incompleteness affects the observed M$_{UV}$ distribution of LBGs (shown as crosses). The fact that there is more contribution from brighter galaxies could, in principle, bias samples toward lower $W_{Ly\alpha}$ when compared to the LFs. We now focus on quantifying all these effects within the framework of our model. 


\begin{figure*}
	\centering
	\includegraphics[width=7in]{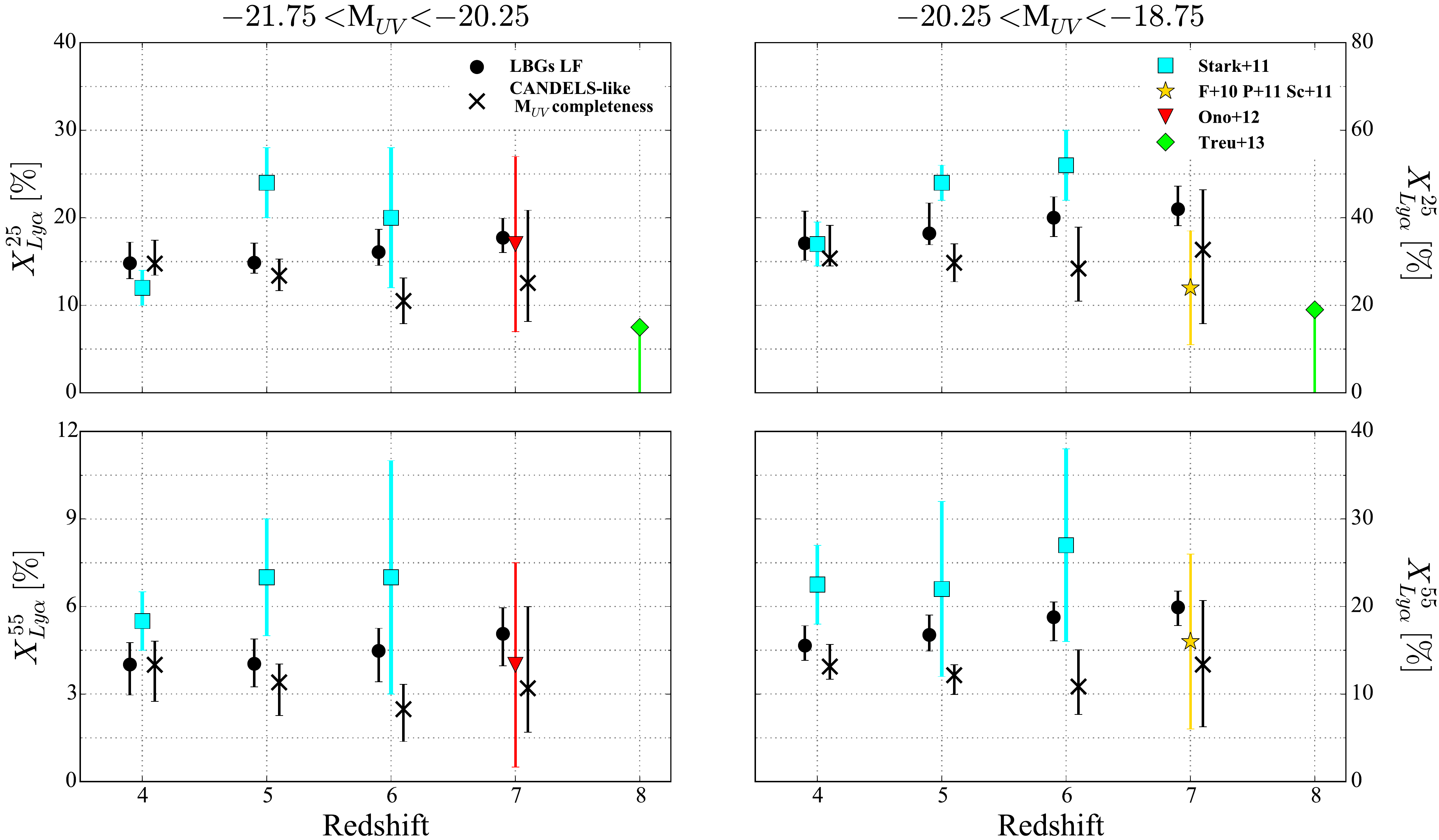}
	\caption{Ly$\alpha$ fraction in brighter ($-21.75<$M$_{UV}<-20.25$; left) and fainter ($-20.25<$M$_{UV}<-18.75$; right) LBGs as a function of redshift. We show the fraction for $W_{Ly\alpha}>25$\AA\ (top) and $W_{Ly\alpha}>55$\AA\ (bottom). Plotted datapoints include observational constraints from \citet{fontana2010}, \citet{pentericci2011}, \citet{stark2011}, \citet{ono2012}, \citet{schenker2012}, and \citet{treu2013}. The datapoint from \citet{ono2012} is a compilation of results from \citet{fontana2010}, \citet{pentericci2011}, \citet{vanzella2011}, \citet{schenker2012}, and \citet{ono2012}. Our $z\sim 4$ characterization of $W_{Ly\alpha}$ in the M$_{UV}-\beta$ plane allows us to predict the Ly$\alpha$ fraction as if the universe was ionized. We show as black circles our constraints based on the M$_{UV}-\beta$ relations from \citet{bouwens2014} and UV LFs from \citet{bouwens2015}. These simulations suggest that the fraction of Ly$\alpha$-emitting LBGs steadily increases up to $z\sim 7$. If we perform the exact same analysis adopting the M$_{UV}$ distribution of CANDELS LBGs (refer to Figure \ref{fig:18}), we recover the fractions shown as crosses. Note how M$_{UV}$ incompleteness artificially lowers the Ly$\alpha$ fraction. This effect must be accounted for when associating Ly$\alpha$ drops with reionization. Keep in mind that our model assumes an exponential $W_{Ly\alpha}$ distribution, which might underestimate the fraction for the tail of the distribution (e.g. $W_{Ly\alpha}>55$\AA; see Figure \ref{fig:4}), especially in populations with high characteristic $W_{Ly\alpha}$ (e.g. $-20.25<$M$_{UV}<-18.75$; see Figure \ref{fig:14}).}	
	\label{fig:19}
\end{figure*}

We simulate the $W_{Ly\alpha}$ distributions for the two sample sets we constructed. To do so, we use the M$_{UV}$ and $\beta$ of each object to draw $W_{Ly\alpha}$ distributions following our sampling of Equations (\ref{Afinal}) and (\ref{W_0final}). This methodology allows us to simulate the $W_{Ly\alpha}$ distribution for the number of galaxies in every redshift subsample, properly taking into account the uncertainties on each M$_{UV}-\beta$ domain and the errors in our modeling. We present the corresponding $4\leqslant z \leqslant 7$ Ly$\alpha$ fractions for such galaxies in Figure \ref{fig:19}. To this end, we select using the thresholds $W_{Ly\alpha}^*$= 25 \AA \ and $W_{Ly\alpha}^*$=55 \AA \ (\citealt{stark2011}). Still, as we note in Section \ref{5.3}, LBGs with M$_{UV}$ distributions dominated by fainter objects can yield higher fractions of Ly$\alpha$. In recent works, characterization has been performed separately in galaxies with M$_{UV}<-20.25$ and M$_{UV}>-20.25$ (\citealt{stark2011,ono2012,schenker2012,tilvi2014,cassata2015}). Following the same approach, we now focus our analysis on samples in the ranges $-21.75<$M$_{UV}<-20.25$ and $-20.25<$M$_{UV}<-18.75$ (Figure \ref{fig:18}), which are well covered by our dataset (see Figure \ref{fig:14}).

We plot in Figure \ref{fig:19} the resulting Ly$\alpha$ fractions. We find the complete fractions (i.e., derived from the LFs; black circles) to be consistent with a steady increase up to $z\sim 7$, which has been previously assumed as baseline for measuring drops in the Ly$\alpha$ fraction (\citealt{stark2010,stark2011,ono2012}). Apart from backing this assumption, our result provides a more robust baseline for comparison. We reiterate that a closer look at Figure \ref{fig:18} reveals the explanation for such trend, at least within our approach. First, the LF of LBGs becomes steeper as redshift increases (\citealt{bouwens2015}). Second, the M$_{UV}-\beta$ relation favors bluer slopes as redshift increases. The combination of both relationships explain an increase in the Ly$\alpha$ fraction as a function of redshift for LBGs. It has already been discussed (\citealt{blanc2011,cassata2015}) how the growth in the Ly$\alpha$ fraction is likely tied to changes in the $f_{esc}$ of Ly$\alpha$ photons, at least for $3<z<5$. This change in the $f_{esc}$ is likely tied to evolution in the dust content and ISM of galaxies, which we are empirically accounting for in this model. Under this picture, Lyman continuum (LyC) photon escape should follow a similar trend. Indeed, \citet{faisst2016} empirically predicts that LyC photon escape fraction increases with redshift at fixed M$_*$. In summary, our results are in strong agreement with the picture that UV photon escape fraction from galaxies tends to decrease with cosmic time.

From the literature, we include in Figure \ref{fig:19} the measurements from \citet{fontana2010}, \citet{pentericci2011}, \citet{stark2011}, \citet{vanzella2011}, \citet{ono2012}, \citet{schenker2012}, and \citet{treu2013}. The overall literature trend is an increase in the Ly$\alpha$ fraction up to $z\sim5$, which flattens at $z\sim6$ and drops toward $z\sim7$. Departures from this increasing regime are typically associated with changes in the transparency of the IGM, with drops toward $z\sim 7$ associated with a more neutral universe (\citealt{stark2010,ono2012}). Even though such inferences might be correct, the contrast between the completeness-corrected (black circles) and -uncorrected (black crosses) fractions in Figure \ref{fig:19} suggests that the driving factor behind these $z\sim7$ drops might be sample M$_{UV}$ incompleteness. Our simulated $z\geqslant 6$ CANDELS LBG samples do not follow the corresponding LF as a result of survey depth (see Figure \ref{fig:18}), leading them to include more luminous objects than a population-representative sample. Therefore, as seen in Figure \ref{fig:19}, drops in the Ly$\alpha$ fraction are induced. As we point out in Section \ref{6.1}, the deepest LBG surveys nowadays (e.g \citealt{bouwens2015,bowler2015,bowler2017,ishigaki2017}) reach $\sim 50\%$ completeness at M$_{UV}\sim$ -18.3, -19, -19.5, and -19.7 for $z\sim$ 4, 5, 6, and 7, respectively. Therefore, Ly$\alpha$ studies of even the deepest LBG surveys are sufficiently incomplete to bias the reported Ly$\alpha$ fractions (see Figures \ref{fig:18} and \ref{fig:19}). Taking into account this effect is essential if we are to use Ly$\alpha$ to further constrain the reionization epoch during the next decade.


It is important to clarify the limitations of our modeling. First, our characterization is based on the whole $W_{Ly\alpha}$ distribution, which may misrepresent the actual fraction of high-$W_{Ly\alpha}$ galaxies. In fact, the analysis we perform in Section \ref{4} reveals that a log-normal profile might be better for reproducing the tail of the distribution (Figure \ref{fig:4}). For example, \citet{zheng2014} find an exponential profile to only adequately represent the low-$W_{Ly\alpha}$ end of the distribution, leading them to model the high-$W_{Ly\alpha}$ tail separately. As a consequence, we encourage the reader to focus most of their analysis on our $W_{Ly\alpha}^*$= 25 \AA \ results (top panel in Figure \ref{fig:19}). Along those lines, we expect our modeling to underestimate the fraction for populations with significantly higher-$W_{Ly\alpha}$ tails. Indeed, our fractions for galaxies with $-20.25<$M$_{UV}<-18.75$ are systematically lower than \citet{stark2011} measurements (see Figure \ref{fig:19}). 

In summary, our results suggest that the Ly$\alpha$ fraction steadily increases up to $z\sim 7$ in an ionized universe. However, drops in this fraction are not necessarily related to changes in the opacity of the IGM, since we find that sample M$_{UV}$ incompleteness can alternatively explain them. In fact, the uncertainties in both literature measurements and our predictions are still large enough to set assertive constraints on the Ly$\alpha$ trends toward $z\sim 7$ (see \citealt{robertson2015}). Our results, therefore, further highlight that Ly$\alpha$ constraints on $z_{re}$ (the redshift at which the fraction of ionized hydrogen is 0.5) are still inconclusive.
	
It is worth discussing whether $z_{re}>7$ is in agreement with observational reionization constraints besides Ly$\alpha$. Recently, the \citet{planck2016b} analysis of CMB anisotropies has yielded $z_{re}\sim 8.2^{+1.0}_{-1.2}$. Systematic detection of Ly$\alpha$-emitting galaxies in the range $z\sim 7-9$ (\citealt{zitrin2015,furusawa2016,stark2017}) still leads to the question whether reionization is already in place at $z\sim 7$. Several studies have also explored Ly$\alpha$ LFs toward $z\sim 9$, although observational limitations require deeper/wider surveys to obtain more meaningful conclusions (\citealt{sobral2009}). Therefore, and since reionization is predicted to happen so rapidly (\citealt{finlator2009}), claims of $z_{re}>7$ are not in tension with measurements of the Ly$\alpha$ fraction at $z\sim 7$. Interestingly enough, it is not even clear whether reionization happens first around fainter, lower density environments (``outside-in," \citealt{kashikawa2006}), or brighter, higher density regions (``inside-out," \citealt{santos2016}). This overall picture suggests that the role of M$_{UV}$ in reionization studies will no longer restrict itself to sample control, but it will also extend to parameterizing property. Therefore, the natural approach for future studies would be to constrain reionization as a smooth function of both redshift and UV luminosity/environment (e.g. \citealt{robertson2013}).

\section{Conclusions}
\label{8}
In this work, we present an exhaustive analysis of Ly$\alpha$ emission at $3<z<4.6$. To this end, we M$_*$-select 625 galaxies from the CANDELS survey, allowing us to study Ly$\alpha$ emission over a diverse and heterogeneous galaxy sample. We conduct spectroscopic observations of our targets with the M2FS, a multi-fiber spectrograph at the Clay 6.5m telescope. We then use a Bayesian approach for proper statistical handling of our results. By means of this framework, we are capable of characterizing Ly$\alpha$ emission in the high-redshift galaxy population. In summary, our conclusions are the following.

1. We present a Bayesian methodology to measure the $W_{Ly\alpha}$ distribution considering the completeness in fluxes measured and the uncertainties in both spectroscopy and photometry. We combine this approach with Monte Carlo simulations for robust modeling of observed trends. Combining all of these features allows us to properly state the significance and uncertainties in our results.

2. We take advantage of the Bayesian framework to give an insight into how to compare multiple $W_{Ly\alpha}$ distribution models from a quantitative standpoint. In this paper, we explore the  extent to which exponential, Gaussian, and log-normal probability distributions can reproduce $W_{Ly\alpha}$ measurements. We conclude that, in our case, an exponential profile is the most adequate. However, as we find in Section \ref{5}, this profile can struggle to reproduce the tail of high-$W_{Ly\alpha}$ populations (e.g. samples with low M$_*$, SFR, UV luminosity, or $\beta$).

3. Our measured $W_{Ly\alpha}$ and $f_{esc}$ strongly anti-correlate with M$_*$. We associate these trends with the higher dust fraction and gas mass in more massive galaxies, boosting the scattering and absorption of Ly$\alpha$ photons. We model both exponential $W_{Ly\alpha}$ distribution parameters ($A, W_0$) using linear relations dependent on M$_*$, and find them both to anti-correlate with M$_*$. Our modeling is also capable of reproducing our observed $W_{Ly\alpha}$ distributions when binning the sample (Figure \ref{fig:7}).

4. We also explore the dependence of Ly$\alpha$ emission on M$_{UV}$. We find $W_{Ly\alpha}$ and $f_{esc}$ to be typically higher for UV faint objects, which have been previously observed in the literature (\citealt{stark2010,schaerer2011}). Since $z\sim 4$ galaxies seem to follow an M$_*-$M$_{UV}$ sequence (\citealt{gonzalez2014}), this result is consistent with our M$_*$ trends. This confirms that the role played by higher dust fraction and gas mass rules $W_{Ly\alpha}$ dependence on UV luminosity. We also observe $W_{Ly\alpha}$ and $f_{esc}$ to anti-correlate with the SFR and UV slope, confirming the qualitative Ly$\alpha$ escape scheme presented.

5. The relatively uniform location of our targets in the M$_{UV}-\beta$ plane allows us to characterize the $W_{Ly\alpha}$ distribution in this space. Once more, we use a Bayesian framework and Monte Carlo simulations to obtain linear representations of exponential $W_{Ly\alpha}$ distribution parameters. Our results suggest that the probability of observing Ly$\alpha$ emission from a galaxy is determined by its dust content, whereas the actual magnitude of $W_{Ly\alpha}$ is mostly ruled by the UV luminosity of the object. This characterization also allows us to simulate $W_{Ly\alpha}$ distributions from M$_{UV}-\beta$ galaxy samples (see below).

6. Using CANDELS, we mimic a $z\sim4$ LBG survey based on B-dropouts. We simulate $W_{Ly\alpha}$ distributions from the LBG sample and verify that the lower probability of measuring high $W_{Ly\alpha}$ when compared to the population is mostly a consequence of detection band limitations. In other words, shallower LBG surveys are bound to include fewer low-M$_*$ galaxies in their samples, decreasing the probability of including high-$W_{Ly\alpha}$ objects. We find color constraints imposed by the LBG technique to also lower the Ly$\alpha$ fraction, although the effect is minor. These results highlight the importance of comparing surveys with similar sensitivities and M$_{UV}$ distributions.

7. Our results on the M$_{UV}-\beta$ plane also allow us to explore the effects $W_{Ly\alpha}$ an line-flux limitations induce on Ly$\alpha$ narrowband surveys. We find $W_{Ly\alpha}$ cuts to bias samples toward low-M$_*$ objects, whereas flux limitations seem to preferentially leave out low-SFR galaxies. These insights contribute to explain the location of LAEs within the main sequence (e.g. some of the differences between \citealt{vargas2014} and \citealt{hagen2014}). 

8. We generate $4\leqslant z \leqslant 7$ LBG samples following reported M$_{UV}-\beta$ relations and LFs from the literature (\citealt{bouwens2014,bouwens2015}). Assuming the same $W_{Ly\alpha}$ dependence on M$_{UV}-\beta$ that we find at $z\sim 4$, we estimate the fraction of Ly$\alpha$ emitting galaxies in LBG samples above an $W_{Ly\alpha}$ threshold. Our findings are consistent with observational measurements in the literature that suggest an increase in the fraction up to $z\sim 6$ (\citealt{stark2010,stark2011,cassata2015}). This result constitutes the first semi-analytical constraint to the Ly$\alpha$ fraction at $z\sim7$, replacing extrapolations of lower redshift regimes.

9. We simulate the $4 \leqslant z \leqslant 7$ Ly$\alpha$ fraction in LBG samples drawing from CANDELS M$_{UV}$ distribution. We conclude that M$_{UV}$ incompleteness can lower the Ly$\alpha$ fraction, reproducing drops at $z\sim 7$. Claims of drops at this redshift need to thoroughly account for this effect if they are to be used to constrain the reionization epoch. This result highlights that reported drops in the Ly$\alpha$ fraction are not inconsistent with $z_{re}>7$. Given the uncertainties in both predictions and measurements, more work is needed to conclude on the actual Ly$\alpha$ trends toward $z\sim 7$ and beyond.


\acknowledgments

We thank the referee for useful feedback that improved the quality of this paper. G.O. was supported by CONICYT, Beca Mag\'ister Nacional 2014, Folio 22140924. G.B. was supported by CONICYT/FONDECYT, Programa de Iniciaci\'on, Folio 11150220. V.G. was supported by CONICYT/FONDECYT, Programa de Iniciaci\'on, Folio 11160832. We also acknowledge the NSF/MRI grant for the M2FS, number AST 0923160. This work is based on observations taken by the 3D-HST Treasury Program (GO 12177 and 12328) with the NASA/ESA HST, which is operated by the Association of Universities for Research in Astronomy, Inc., under NASA contract NAS5-26555. This research has made use of the NASA/IPAC Extragalactic Database (NED), which is operated by the Jet Propulsion Laboratory, California Institute of Technology, under contract with the National Aeronautics and Space Administration. This paper includes data gathered with the 6.5 meter Magellan Telescopes located at Las Campanas Observatory, Chile.
\bibliography{mybib}
\end{document}